\documentclass[onecolumn]{article}
\usepackage{amsmath,amssymb}

\input xy
\xyoption{all}
\xyoption{arc}
\usepackage[curve]{xy}
\xyoption{poly}
\usepackage{hyperref}
\begin{document}
\large  
\centerline{\Large {\bf Cosmological Coincidence and Dark Mass}}
\centerline{\Large {\bf Problems in Einstein Universe and Friedman}}
\centerline{\Large {\bf Dust Universe with Einstein's Lambda}}
\centerline{\Large {\bf Quantum Cosmology Dark Energy}}
\centerline{\Large {\bf Schr\"odinger Wave Motion}}
\vskip0.5cm
\centerline{James G. Gilson\quad  j.g.gilson@qmul.ac.uk}
\centerline{School of Mathematical Sciences}
\centerline{Queen Mary University of London}
\centerline{Mile End Road London E14NS}
\vskip 0.7cm 
\centerline{May 20th  2007}
\vskip 0.7cm  
\large \centerline{\bf Abstract}
\vskip 0.7cm
In this paper, it is shown that the cosmological model that was introduced
in a sequence of three earlier papers under the title  {\it A Dust Universe Solution to the Dark Energy Problem\/} can be used to analyse and solve the {\it Cosmological Coincidence Problem\/}. The generic coincidence problem that appears in the original Einstein universe model is shown to arise from a misunderstanding about the magnitude of dark energy density and the epoch time governing the appearance of the integer relation between dark energy and normal energy density. The solution to the generic case then clearly points to the source of the time coincidence integer problem in the Friedman dust universe model. It is then possible to eliminate this coincidence by removing a degeneracy between different measurement epoch times. In this paper's first appendix, a fundamental time dependent relation between dark mass and dark energy is derived with suggestions how this relation could explain cosmological voids and the clumping of dark mass to become visible matter. In this paper's second appendix, it is shown that that dark energy is a conserved with time substance that is everywhere and for all time permeable to the dark mass and visible mass of which the contracting or expanding universe is composed. The last two appendices involve detailed studies of cosmology, quantum dark energy related issues. There are more detailed abstracts given with all four appendices. 
\vspace{0.4cm}
\vskip 0.5cm 
\centerline{Keywords: Dust Universe, Dark Energy, Dark Mass, Friedman Equations,}
\centerline{Zero-Point Energy, Cosmological Voids, Coincidence Problem}
\vskip 0.2cm
\centerline{PACS Nos.: 98.80.-k, 98.80.Es, 98.80.Jk, 98.80.Qc}
\section{Introduction}
\setcounter{equation}{0}
\label{sec-intr}
The work to be described in this paper is an application of the cosmological model introduced in the papers {\it A Dust Universe Solution to the Dark Energy Problem\/} \cite{45:gil}, {\it Existence of Negative Gravity Material. Identification of Dark Energy\/} \cite{46:gil} and {\it Thermodynamics of a Dust Universe\/} \cite{56:gil}. The conclusions arrived at in those papers was that the dark energy {\it substance\/} is physical material with a positive density, as is usual, but with a negative gravity, -G,  characteristic and is twice as abundant as has usually been considered to be the case.  References to equations in those papers will be prefaced with the letter $A$, $B$ and $C$ respectively. The work in $A$, $B$ and $C$, and the application here have origins in the studies of Einstein's general relativity in the Friedman equations context to be found in references (\cite{03:rind},\cite{43:nar},\cite{42:gil},\cite{41:gil},\cite{40:gil},\cite{39:gil},\cite{04:gil},\cite{45:gil}) and similarly motivated work in references (\cite{10:kil},\cite{09:bas},\cite{08:kil},\cite{07:edd},\cite{05:gil}) and 
(\cite{19:gil},\cite{28:dir},\cite{32:gil},\cite{33:mcp},\cite{07:edd},\cite{47:lem},\cite{44:berr}).
Other useful sources of information are (\cite{3:mis},\cite{44:berr},\cite{53:pap},\cite{49:man},\cite{52:ham},\cite{54:riz}) with the measurement essentials coming from references (\cite{01:kmo},\cite{02:rie},\cite{18:moh}).  Further references will be mentioned as necessary.  
The  application of the cosmological model introduced in the papers $A$ \cite{45:gil}, $B$,\cite{46:gil} and $C$ \cite{56:gil} is to the extensively discussed and analysed {\it Cosmological Coincidence Problem\/}. This problem arose from Einstein's time static cosmology model derived from his theory of general relativity. The Einstein first model is easily obtained from the Friedman equations (\ref{S1}) and (\ref{S2}) with the positively valued $\Lambda >0$ term that he introduced to prevent his theoretical universe from collapsing under the gravitational pull of its material contents, 
\begin{eqnarray}
8\pi G \rho r^2/3 & = & {\dot r}^2 +(k - \Lambda r^2/3) c^2\label{S1}\\
-8 \pi GP r/c^2 & = & 2 \ddot r + {\dot r}^2/r +(k/r -\Lambda r) c^2 .\label{S2}
\end{eqnarray}
Einstein's preferential universe was of the closed variety which involves the curvature parameter being unity, $k=1$ and with a positively valued $\Lambda$, we have,
\begin{eqnarray}
8\pi G \rho r^2/3 & = & {\dot r}^2 +(1 - \Lambda r^2/3) c^2\label{S3}\\
-8 \pi GP r/c^2 & = & 2 \ddot r + {\dot r}^2/r +(1/r -\Lambda r) c^2 .\label{S4}
\end{eqnarray}
To get a static universe from these equations that holds for some finite time interval we have to impose the non expansion condition, $v =\dot r =0$, together with the none acceleration condition $a=\dot v= \ddot r =0$ and if, additionally,  we choose the dust universe condition, $P=0$, we get (\ref{S5}) and (\ref{S6}). 
\begin{eqnarray}
8\pi G \rho r^2/3 & = &  (1 - \Lambda r^2/3) c^2\label{S5}\\
0 & = & (1/r -\Lambda r) c^2. \label{S6}
\end{eqnarray}
Einstein identified his cosmological constant $\Lambda$ as arising from a density of {\it dark energy\/} in the vacuum,
 $\rho _\Lambda = \Lambda c^2/(8\pi G)$, so that equations (\ref{S5}) and (\ref{S6}) could be put into the forms (\ref{S7}) and (\ref{S8}) with the radius of the Einstein universe given by (\ref{S8.1}),
\begin{eqnarray}
8\pi G (\rho +  \rho _\Lambda )& = & 3 c^2/r_E^2\label{S7}\\
8\pi G (\rho_\Lambda) & = &  c^2/r_E^2 \label{S8}\\
r_E &=& \Lambda ^{-1/2}.\label{S8.1}
\end{eqnarray}
From (\ref{S7}) and (\ref{S8}) it follows that
\begin{eqnarray}
8\pi G \rho & = & 2c^2/r_E^2\label{S9}\\
\rho_\Lambda & = & \rho /2\label{S10}\\
\rho & = & 2\rho_\Lambda = \rho_\Lambda^\dagger .\label{S10.1}
\end{eqnarray}
Equation (\ref{S10.1}) is the {\it generic\/} version of the so called {\it cosmological coincidence problem\/}. I think that Einstein would not have recognised the relationship between $\rho$ and $\rho_\Lambda$ at (\ref{S10.1}) as a problem in the early years after discovering it. He probably thought that the $2$ factor was interesting and needed explaining but did not see it as a problem. In those early years he was convinced the universe was a time static entity and had no vision of the possibility that the relation might have a different coefficient from the integer $2$ which could come about by the now recognised and accepted expansion process.  Only after expansion was accepted does the question following arise.  If at time now equation (\ref{S10.1}) holds in an expanding universe of decreasing density $\rho$ with time and with $\rho _\Lambda$ an absolute constant, is it not an {\it extraordinary\/} coincidence that at time now the coefficient in (\ref{S10.1}) is exactly the integer {\it 2\/}? Clearly the significance of the factor $2$ must be seen against the likely possible values of $\rho$ which probably varies from $\infty$ to $0$ with $\rho _\Lambda$ remaining constant over the whole positive life time history of the universe.
Einstein's generic cosmological coincidence problem is completely resolved by the cosmological model introduced in references $A$ \cite{45:gil}, $B$,\cite{46:gil} and $C$ \cite{56:gil} as I shall next explain. However, there is one {\it important\/} reservation about this claim that will be discussed in the next section. I call this first cosmological coincidence {\it critical\/} because it involves the integer {\it point value\/} number, $2$,  which would have zero probability of occurring in any finite time ranged variable quantity. Such coincidences need to be explained in any structure.
The model introduced in those papers reveals the true nature of dark energy material and that is the clue to resolving the generic coincidence problem.
One conclusion from those papers was that the dark energy density, contrary to Einstein's identification, should be theoretically and physically measured as $\rho_\Lambda^\dagger$ (\ref{S10.1}) rather than as $\rho_\Lambda$. The second conclusion from those papers was that dark energy has positive mass density but is characterised by carrying a negative gravitational value of the gravitational constant, $-|G|$. Thus equation (\ref{S10.1}) achieves Einstein's purpose of stopping the gravitation collapse of the universe by choosing conditions such that the positive mass material, $\rho +\rho _\Lambda^\dagger$, within the universe is {\it gravitationally\/} neutral, $G\rho +(-G)\rho _\Lambda^\dagger =0$. Thus although that could have happened some time or other it would not necessarily hold for ever as in a constant universe or indeed occur at the time now. The model I am suggesting is a flat universe with, $k=0$, and the actual time when such conditions apply is denoted by $t_c$ and can be calculated. At that time $v(t_c)\not =0$ contrary to the what is implied in the Einstein universe where $v =0$ given above. The time $t_c$ is the important time greatly in the past and recognised recently by astronomers when the acceleration of the universe changes through zero from negative to positive or when dark energy takes over from normal mass energy.    
The {\it critical\/}  coincidence in the generic Einstein universe is completely resolved by the conceptual aspects of the Friedman dust universe that I have been proposing. This reinterpreted old and modified model which is closely related to an early Lema\^itre model has a structure that  has identified  the cause of the Einstein critical coincidence. The nature of this coincidence can be described as, {\it mistaking the Einstein radius for a possible constant present time radius\/}. This mistake is completely excusable on the grounds that Einstein did not recognise that the universe radius was in truth a variable with time quantity and he was completely unaware that at some time in the past the dark energy density as he defined it was exactly half the normal mass density.  
The explanation of the root cause of the critical Einstein coincidence can be used to identified the cause of another critical time coincidence between the present time $t^\dagger$ and time $t_c$, $ t^\dagger = 2 t_c$, in the Friedman dust universe. This will be explained in the next section.
\section{Coincidence in Friedman Dust Universe}
\setcounter{equation}{0}
\label{sec-cfdu}
The coincidence in the Friedman dust universe model involves, $t^\dagger$, the time now and, $t_c$, the time when the universe changed from deceleration to acceleration.
\begin{eqnarray}
t^\dagger = 2 t_c.\label{S17}
\end{eqnarray}
This equation involves again the exact numerical integer value, $2$. This is clearly $critical$ because if two events over time are so related, then there must be some physical explanation because the probability of two such time-point events on any finite time line range is zero.
The generic Einstein coincidence was critical in the same sense. This coincidence seems obviously related to the generic Einstein coincidence which suggests it is also totally explainable. The reservation I mentioned earlier is that you might see it as ironic that a model with a coincidence can completely solve the coincidence in an earlier model. This can be explained by the fact that theoretical structures involve patterns of abstract symbols as one aspect and numerical constants as another aspect when they are applied to physical situations. The new model is correct in the first aspect but in the second aspect, the numerical values have not all been associated with the measurement time, $t^\dagger$, but rather some with a conceptual time, $t_0$, the time that would be associated with the centre of the values given by the astronomical measurements. There is some subtlety in this situation because in this model, it seemed that $t^\dagger$ should be equal to $t_0$. However, this equality created the degeneracy that led to the coincidence. It can all be resolved by using the formula for Hubble's constant, the formula for the radius and the formula for the constant C,   
\begin{eqnarray}
H(t) &=& (c/R_\Lambda) \coth(3 c t/(2 R_\Lambda )) \label{S18}\\
r(t) &=& b\sinh ^{2/3}(3 c t/(2 R_\Lambda)) \label{S19}\\
C &=&  \Omega _{M,0} H^2(t_0)r^3(t_0). \label{S20}
\end{eqnarray}
These expressions involve the numerical parameter, $R_\Lambda$. It is necessary to find the correct value for this parameter that is to be associated with these formulae. To make this step, we need the astronomical measurements of the $\Omega s$.
The accelerating universe astronomical observational workers \cite{01:kmo} give measured values of the three $\Omega s$, and $w_\Lambda$ to be
\begin{eqnarray}
\Omega_{M,0} &=&8\pi G\rho_0/(3 H_0^2)=0.25^{+0.07}_{-0.06}\label{9}\\
\Omega_{\Lambda,0} &=& \Lambda c^2/(3 H_0^2)=0.75^{+0.06}_{-0.07}\label{10}\\
\Omega_{k,0} &=& -kc^2/(r_0^2 H_0^2) =0,\ \Rightarrow k= 0\label{11}\\
\omega_\Lambda &=& P_\Lambda/(c^2 \rho _\Lambda) = -1\pm\approx 0.3.\label{11.1}
\end{eqnarray}
From these equations assumed to hold at a conceptual time, $t_0$, when the universe passes through the centre value of the measurement ranges, we get the formulae,
\begin{eqnarray}
t_0&=& (2R_\Lambda/(3c))\cosh^{-1}(2)\label{12}\\
R_\Lambda &=& 3ct_0/(2 \cosh^{-1}(2))\label{13}\\
t_c &=& (2R_\Lambda/(3c)\coth^{-1}(3^{1/2})\label{14}\\
t_0/t_c &=& \cosh^{-1}(2)/\coth^{-1}(3^{1/2})=2.\label{15}
\end{eqnarray}
Having found $R_\Lambda$ in terms of $t_0$ this value of $R_\Lambda$ can be substituted into the formula for Hubble's constant, (\ref{S18}),  to find the value of the {\it time now\/}, $t^\dagger$. 
\begin{eqnarray}
H(t^\dagger) &=& (c/R_\Lambda) \coth(3 c t^\dagger /(2 R_\Lambda )) \label{16}\\
t^\dagger &=& (2R_\Lambda /(3 c) )\coth^{-1} (R_\Lambda H^\dagger/c)\label{17}\\
&=& \left(   \frac{t_0}{\cosh^{-1}(2)}  \right)\left(  \coth^{-1}\left(  \frac{3t_0H^\dagger }{2\cosh^{-1}(2)}\right)\right)\label{18}\\
&=& \left(   \frac{2t_c}{\cosh^{-1}(2)}  \right)\left(  \coth^{-1}\left(  \frac{6t_c H^\dagger }{2\cosh^{-1}(2)}\right)\right),\label{19}
\end{eqnarray}
where $H^\dagger= H(t^\dagger)$ is the present day measured value of Hubble's constant.
Equations (\ref{18}) or (\ref{19}) is essentially the solution to the coincidence problem. If we write (\ref{19}) in the form
\begin{eqnarray}
t^\dagger /t_c &=& \left(\frac{2}{\cosh^{-1}(2)}  \right)\left(\coth^{-1}\left(\frac{6t_c H^\dagger}
{2\cosh^{-1}(2)}\right)\right)\label{20}\\
t^\dagger /t_c &=& 2f(2 t_c),\label{21}
\end{eqnarray}
where $f(2t_c)$  gives the deviation of the ratio $t^\dagger /t_0$ from the value unity and removes the degeneracy. Expressed in another way it is the multiplicative function that breaks the coincidence at (\ref{15}) and converts the integer $2$ to a much less notable non integral value. However, we can give the formulae (\ref{20}) and (\ref{21}) together an interpretation in terms of the uncertainties of the measurement process.
This is achieved by defining the measurement {\it deviation\/} function $d_{meas}(t_0)$ as follows,
\begin{eqnarray}
d_{meas}(t_0)&=& t^\dagger /t_0  - f(t_0) \label{22}\\
f(t_0) &=& \left(\frac{1}{\cosh^{-1}(2)}  \right)\left(\coth^{-1}\left(  \frac{3t_0 H^\dagger }{2\cosh^{-1}(2)}\right)\right) .\label{23}
\end{eqnarray}
The function (\ref{22}) is a dimensionless measure of how much the central $\Omega$ values from astronomy assumed to have occurred at $t_0$ differ from the time now measurement from the Hubble variable quantity $H(t^\dagger)$ taken at time now, $t^\dagger$. It is sufficient to assume that the event at $t_0$ is still yet to occur, $t_0 > t^\dagger$, then we see that the function $d_{meas}$ passes through zero when the full degeneracy holds at $t_0 = t^\dagger$ and it has a maximum at $t_0\approx 0.643\times 10^{18} s$ when $t^\dagger$ and $t_0 $ assume the approximate maximum deviation, $0.17$.
When $t_0=0.643\times 10^{18}$, $t^\dagger$ can be assumed constant at the coincidence value $4.34467 \times 10^{18}$  so that the maximum deviation times ratio is $t^\dagger /t_0 \approx 0.43467 / 0.643 \approx 0.6757$ or
\begin{eqnarray}
t^\dagger =0.6757t_0.\label{24}
\end{eqnarray}
It follows that $t^\dagger$, the time now value, can vary from $t_0$ down to a value of $t^\dagger \approx 0.6757t_0 =1.3514t_c$. Thus the coincidence is decisively removed with $t^\dagger \not = t_0 = 2 t_c$.
\section{Conclusions}
It has been shown that the generic Einstein coincidence problem can be resolved in terms of a correction in the value of the density he associated with his cosmological constant $\Lambda$ and a rethink about the significance of the radius of his model. This solution then points clearly to resolution of the coincidence in the recent dust universe model as essentially the same concepts
are involved. The conceptual centre  $\Omega$ value measurements from the astronomers can not necessarily be assumed to occur at exactly the same epoch time $t_0$ as the measurement of the value of the Hubble constant at epoch time now, $t^\dagger$. The usually assumed degeneracy $t_0 = t^\dagger$ can be removed to find the true range of values within which $t^\dagger$ has to reside so that the integer $2$ aspect of the same degeneracy $t^\dagger = 2t_c$ sees the $2$ replaced with a less mysterious non integer. The time $t_c$ is when the expansion acceleration changes from negative to positive.
\vskip 0.5 cm   
\centerline{\bf Acknowledgements}
\vskip 0.5 cm   
I am greatly indebted to Professors Clive Kilmister and Wolfgang Rindler for
help, encouragement and inspiration over many years.
\vskip 0.5 cm
\section{Appendix 1}
\setcounter{equation}{0}
\label{sec-abs1} 
\vskip 0.5 cm
\centerline{Fundamental Dark Mass, Dark Energy Time}
\centerline{Relation in a Friedman Dust Universe}
\centerline{and in a Newtonian Universe}
\centerline{with Einstein's Lambda}
\vskip 0.5 cm 
\centerline{\bf Appendix 1 Abstract}
\vskip 0.7cm
In this appendix, it is shown that the cosmological model that was introduced
in a sequence of three earlier papers under the title  {\it A Dust Universe Solution to the Dark Energy Problem\/} can be used to recognise a fundamental time dependent relational process between dark energy and dark mass. It is shown that the formalism for this process can also be obtained from Newtonian gravitational theory with only the additional assumption that Newtonian space contains a constant universal dark energy density distribution dependant on Einstein's Lambda, $\Lambda$. It thus seems that the process is independent of general relativity and applies in more contexts than just the expansion of the entire universe. It is suggested that the process can be thought of as a local space and time packaging for dark mass going through part transmutations into locally condensed visible material. The process involves a contracting and then expanding sphere of conserved dark matter. At two stages in the process at special times before and after a singularity at time zero, the spherical package goes through a condition of gravitational neutrality of very low mass density which could be identified as cosmological voids. The process is an embodiment of the principle of equivalence.
In the next section a relation between Dark Mass and Dark Energy over epoch time is deduced and analysed.
\section{Cosmological Vacuum Polarisation}
\setcounter{equation}{0}
\label{sec-cvp} 
Consider the result for gravitational vacuum polarisation derived in paper (D)
\begin{eqnarray}
G\rho_\Lambda &=& G_{-} \Gamma_B(t)   + G_{+} \Delta _B(t)\label{B0}\\
0& =& G_{-} \Gamma_Z(t)   + G_{+} \Delta _Z(t),\label{B0.1}
\end{eqnarray}
where $ G_{-}=-G$ and $ G_{+} =G$. The upper case Greek functions $\Gamma_B(t)$, $\Delta _B(t)$, $\Gamma_Z(t)$ and $\Delta _Z(t)$  are defined from the equations of state for $\Delta$ and $\Gamma$ substances which together are assumed to form all the time conserved material of the universe,
\begin{eqnarray}
P_{\Delta B}/c^2= \rho_{\Delta B,\nu_c}(t)\omega_\Delta (t)&=&\Delta_B(t) \label{B1}\\
P_{\Gamma B}/c^2=\ \rho_{\Gamma B,\nu_c}(t)\omega_\Gamma (t)&=&\ \Gamma_B(t)\label{B2}\\
P_{\Delta Z}/c^2=\rho_{\Delta Z,\nu_c}(t)\omega_\Delta (t)&=&\Delta_Z(t) \label{B3}\\
P_{\Gamma Z}/c^2=\ \rho_{\Gamma Z,\nu_c}(t)\omega_\Gamma (t)&=&\ \Gamma_Z(t).\label{B4}
\end{eqnarray}
The $Z$ subscript above denotes zero-point values.
Let us now consider the Einstein cosmological constant, $\Lambda$, in relation to the Friedman equations,
\begin{eqnarray}
8\pi G \rho r^2/3 & = & {\dot r}^2 +(k - \Lambda r^2/3) c^2\label{B5}\\
-8 \pi GP r/c^2 & = & 2 \ddot r + {\dot r}^2/r +(k/r -\Lambda r) c^2 .\label{B6}
\end{eqnarray}
Einstein introduced a physical explanation for his $\Lambda$ term by associating it with a density of what is nowadays called {\it dark energy\/} in the form of an additional {\it mass\/} density, $\rho_\Lambda$, where  $\rho _\Lambda =\Lambda c^2/(8\pi G)$. Thus with this density the Friedman equations can be written with the Hubble function of epoch time $H(t)$ as,
\begin{eqnarray} 
8\pi G \rho r^2/3 & = & {\dot r}^2 +(k - 8\pi G\rho_\Lambda r^2/3)c^2  \label{B7}\\
-8 \pi GP r/c^2 & = & 2 \ddot r + {\dot r}^2/r +(k c^2 /r -8\pi G\rho_\Lambda r) \label{B8}\\
H(t) &=& \dot r(t)/r(t)= (c/(R_\Lambda))\coth (3ct/(2R_\lambda)).\label{B9}
\end{eqnarray}
Thus the first friedman equation can be expressed as
\begin{eqnarray} 
8\pi G (\rho +\rho_\Lambda)/3 & = & H^2 (t) +(k c^2 /r^2) \label{B10}\\
8\pi G \rho^T_E &=&3( H^2 (t) +k c^2 /r^2) \label{B11}\\
\rho^T_E&=&\rho +\rho_\Lambda,\label{B12}
\end{eqnarray}
where $\rho^T_E $ is the total density for mass at points within the boundary of the universe as perceived by Einstein.
Rearranging the first Friedman equation, we have
\begin{eqnarray} 
8\pi G (\rho +\rho_\Lambda) -3(k c^2 /r^2) & = & 3H^2 (t) \label{B13}\\
\frac{8\pi G \rho }{3H^2 (t)}+\frac{8\pi G\rho_\Lambda }{3H^2 (t)} -\frac{k c^2 }{r^2H^2 (t)}& =&1.\label{B14}
\end{eqnarray}
The three Omegas which the astronomers use to display their measurements are defined using the three terms on the left hand side of (\ref{B14}) according to which they have to add up to {\it unity\/},
\begin{eqnarray} 
\Omega_M(t)&=&8\pi G \rho /(3H^2 (t)) \label{B15}\\
\Omega_\Lambda(t)&=&8\pi G\rho_\Lambda /(3H^2 (t)) \label{B16}\\
\Omega _k(t)&=&-k c^2 /(r^2H^2 (t))\label{B17}\\
\Omega_M(t) + \Omega_\Lambda(t) + \Omega_k(t) &=&1.\label{B18}
\end{eqnarray}
There is a very strong case (A,B,C,D,E) for identifying the dark energy mass density that should account for Einstein's constant $\Lambda$ term as given by twice the density introduced by Einstein, 
\begin{eqnarray} 
\rho^\dagger_\Lambda = 2\rho_\Lambda\label{B19}\\
\rho^{T\dagger} = \rho + \rho^\dagger_\Lambda\label{B20}
\end{eqnarray}
and this implies the formula (\ref{B20}) for the total amount of physical mass density within the boundaries of the spherical universe in contrast with (\ref{B12}). Thus equation (\ref{B13}) should be replaced by
\begin{eqnarray} 
8\pi G (\rho +\rho^\dagger_\Lambda) -3(k c^2 /r^2) & = & 3H^2 (t) + 8\pi G \rho_\Lambda\label{B21}\\
\frac{8\pi G \rho }{3H^2 (t) + c^2\Lambda}+\frac{8\pi G\rho^\dagger_\Lambda }{3H^2 (t) + c^2\Lambda } &-&\frac{3k c^2 }{r^2(3 H^2 (t) + c^2\Lambda)} =1.\label{B22}
\end{eqnarray}
Thus we now have three new Omegas
\begin{eqnarray} 
\Omega^\dagger _M(t)&=&8\pi G \rho /(3H^2 (t) + c^2\Lambda) \label{B23}\\
\Omega^\dagger _\Lambda(t)&=&8\pi G\rho^\dagger_\Lambda /(3H^2 (t) + c^2\Lambda) \label{B24}\\
\Omega^\dagger _k(t)&=&-k 3c^2 /( r^2(3H^2 (t) + c^2\Lambda))\label{B25}\\
\Omega^\dagger _M(t) + \Omega^\dagger _\Lambda(t) + \Omega^\dagger _k(t) &=&1.\label{B26}
\end{eqnarray}
Here I shall be mostly concerned with the flat space case $k=0$ so that the two possible and equivalent sets of Omegas  satisfy the relations 
\begin{eqnarray} 
\Omega_M(t) + \Omega_\Lambda(t)  &=&1\label{B27}\\
\Omega^\dagger _M(t) + \Omega^\dagger _\Lambda(t) &=&1.\label{B28}
\end{eqnarray}
Inspection of the formulae for $H(t)$, $\Omega _M (t)$ and $\Omega _\Lambda (t)$ shows that $\Omega_\Lambda (t)$ varies between $0$ and $1$ as $t$ varies between $0$ and $\infty$ and consequently from (\ref{B27}), $\Omega_M(t)$ varies between $1$ and $0$. It follows that there will be a time when
\begin{eqnarray} 
\Omega_M(t_0)&=&1/4\label{B29}\\
\Omega_\Lambda(t_0) &=&3/4\label{B30}
\end{eqnarray}
and this event will happen regardless of any measurements. I have assumed that the epoch time of this event in the history of the universe is given by $t_0$. Thus the usual use of the subscript $0$ to denote {\it time now\/} has been abandoned and {\it time now\/} will in future be denoted by $t^\dagger$. The corresponding and more realistic time $t_0$ relation between non-dark energy materials and dark energy will with a simple calculation be represented in terms of the dagger Omegas by
\begin{eqnarray} 
\Omega^\dagger_M(t_0)&=&1/7\label{B31}\\
\Omega^\dagger_\Lambda(t_0) &=&6/7.\label{B32}
\end{eqnarray}
This implies that about $85.7\%$ of the universe mass is dark energy rather than the usually assumed $75\%$, a {\it substantially\/} changed assessment. If this assessment of the percentage of dark energy to conserved mass is accepted, it will also have some effect on the amount of {\it visible mass\/} assumed to be present within the total mass of the universe. The ratio dark mass to visible mass is often taken to be $4$ to $1$. Thus the percentage of dark mass in the universe according to (\ref{B31}) and (\ref{B32}) would become reduced to $20\times (4/7)\% \approx 11.44\%$. The total non-visible mass would then be $85.7\% + 11.44 \% \approx 97.14\%$ leaving us with being able to see just about $2.86\%$ of the total mass. If it is taken that we know nothing about the dark elements, as is often suggested, then our actual knowledge of the universe is mass wise abysmal. However, fortunately it is not true that we have {\it no\/} knowledge of the dark elements. We do have indirect knowledge of these aspects.
The theory associated with this model give a definite relation between dark energy and dark mass this relation can be read off from the gravitation polarisation equations (\ref{B0}, \ref{B0.1}) repeated next
\begin{eqnarray}
G\rho_\Lambda &=& G_{-} \Gamma_B(t)   + G_{+} \Delta _B(t)\label{B33}\\
0& =& G_{-} \Gamma_Z(t)   + G_{+} \Delta _Z(t)\label{B34}\\
\rho (t)&=& \rho_{\Delta,\nu_c} +\rho_{\Gamma,\nu_c}.\label{B35} 
\end{eqnarray}
The third equation above expresses the total time conserved density $\rho (t)$ in terms of the $CMB$ mass density, $\rho_{\Gamma,\nu_c}$ , and the rest of the universe mass density $\rho_{\Delta,\nu_c}$. The $\nu_c$ subscript indicates that zero point energies are included in these terms. The second equation above defines the zero-point energy of the dark energy as being zero, effectively defining energy zero for this cosmology theory. The total energy density for this model equation (\ref{B20}) can thus be written as (\ref{B39})
\begin{eqnarray} 
\rho^\dagger_\Lambda &=& 2\rho_\Lambda\label{B36}\\
\rho^{T\dagger}(t) &=& \rho (t) + 2 \rho_\Lambda\label{B37}\\
\rho^{T\dagger}(t) &=& \rho_{\Delta,\nu_c} +\rho_{\Gamma,\nu_c}+ 2 (\Delta _B(t)- \Gamma_B(t))\label{B38}\\
\rho^{T\dagger}(t) &=& \rho_{\Delta,\nu_c}+ 2 \Delta _B(t)+\rho_{\Gamma,\nu_c}-2\Gamma_B(t)\label{B39}\\
\rho^{T\dagger}(t) &=& \tilde\rho_{\Delta,\nu_c}+\tilde\rho_{\Gamma,\nu_c} \label{B40}\\
\tilde\rho_{\Delta,\nu_c}&=&\rho_{\Delta,\nu_c}+2\Delta_B(t)\label{B41}\\
\tilde\rho_{\Gamma,\nu_c}&=&\rho_{\Gamma,\nu_c}-2\Gamma_B(t).\label{B42}
\end{eqnarray}
The tilde versions of the basic two densities are the resultants of a gravitational vacuum polarisation process in which the basic $\Gamma$ and $\Delta$ densities induce, via their pressures and coexistence, the two polarisation densities $2\Gamma_B (t)$ and $2\Delta _B (t)$ which together represent the dark energy density $\rho _\Lambda$, equation (\ref{B33}).  This process takes place through the equations of motion of the two components. Thus from this point of view dark energy within the universe boundary is a vacuum polarisation consequence of the of the existence of the basic $\Gamma$ and $\Delta$ fields in interaction under general relativity. The dark energy density also exists outside the universe boundary but in an un-polarised condition. Thus the polarisation within the universe is constrained by the constant value that exists everywhere. To examine the weight of this gravitational vacuum polarisation on the none polarised fields separately at time $t^\dagger$ using the numerical results from (A,B,C)
\begin{eqnarray}
2\omega_\Delta (t^\dagger) &\approx& 6\label{B43}\\
2\omega_\Gamma (t^\dagger)&=& 2/3\label{B44}
\end{eqnarray}
they must be expressed in terms off the none polarised fields as in (\ref{B45}) and (\ref{B46})
\begin{eqnarray}
2\Delta_B(t^\dagger) &\approx& 6\rho_{\Delta B,\nu_c}( t^\dagger)\label{B45}\\
2\Gamma_B(t^\dagger) &=& (2/3)\rho_{\Gamma B,\nu_c}( t^\dagger)\label{B46}\\
\rho_{\Delta B,\nu_c}( t^\dagger) &\approx & ({10^4}/1.9)\rho_{\Gamma B,\nu_c}( t^\dagger)\label{B47}\\
\rho_{\Gamma B,\nu_c}( t^\dagger) &\approx& 1.9\times 10^{-4}\rho_{\Delta B,\nu_c}(t^\dagger)\label{B48}\\
\rho_\Lambda &=& \Lambda c^2/(8 \pi G) \approx 7.3 \times 10^{-27}\label{B48.1}\\
\rho_{\Gamma B,\nu_c}( t^\dagger) &=& a T^4(t^\dagger) \approx 4.66 \times 10^{-31}.\label{B48.2}
\end{eqnarray}
The relation (\ref{B47}) also comes from (A,B,C).
Thus we can express the positively weighted $\Delta_B$ and negative weighted $\Gamma_B$ vacuum polarisation density poles as
\begin{eqnarray}
2\Delta_B(t^\dagger) &\approx &(6 \times 10^4/1.9)\rho_{\Gamma B,\nu_c}( t^\dagger)\label{B49}\\
2\Gamma_B(t^\dagger) &\approx & (2/3) 1.9\times 10^{-4}\rho_{\Delta B,\nu_c}( t^\dagger)\label{B50}\\
2\Delta_B(t^\dagger) &\approx& 3\times 10^4\rho_{\Gamma B,\nu_c}( t^\dagger)\label{B51}\\ 
2\Gamma_B(t^\dagger) &\approx& 1.26\times 10^{-4}\rho_{\Delta B,\nu_c}( t^\dagger).\label{B52} 
\end{eqnarray}
Returning to the gravitational vacuum polarisation equation (\ref{B0}) repeated here for convenience,
\begin{eqnarray}
G\rho_\Lambda &=& G_{-} \Gamma_B(t)   + G_{+} \Delta _B(t)\label{B53}\\
0& =& G_{-} \Gamma_Z(t)   + G_{+} \Delta _Z(t),\label{B54}
\end{eqnarray}
we can do a spot numerical check using the values above and without the G factor as follows
\begin{eqnarray}
7.3\times 10^{-27}&\approx&\rho_\Lambda =  \Delta _B(t^\dagger) - \Gamma_B(t^\dagger)\label{B55}\\
& =& \rho_\Delta\omega_\Delta -\rho_\Gamma \omega_\Gamma\label{B56}\\
& \approx& (3\times 10^{4} -(1/3))\rho_\Gamma\label{B57}\\
& \approx& (3\times 10^{4})\rho_\Gamma\label{B58}\\
& \approx& (3\times 10^{4})\times 4.66\times 10^{-31}\label{B59}\\
& \approx& (13.98/1,9)\times 10^{-27}\label{B60}\\
& \approx& 7.3\times 10^{-27}.\label{B61}
\end{eqnarray}
This is just a rough check that does give a good though approximate result while showing that the induced $\Delta$ and induced $\Gamma$ fields in the form of a difference are the {\it source\/} of the dark energy density within the universes boundaries. At step (\ref{B57}), the $-1/3$ term from the $\Gamma$ field is abandoned because it contributes negligibly in relation to the $10^{4}$ from the $\Delta$ term. However, at step (\ref{B58}) the $\Gamma$ field only appears to be a main contributor  because it occurs as multiplicatively weighted by the $\Delta$ factor, $10^{4}$. As the $\Delta$ field is all the conserved universe field density less the $CMB$ the induced delta field $\Delta$ is all the induced conserved density universe field less the induced $CMB$ field. The $\Delta$ field includes the so called {\it dark matter\/} as its major contributor of about $80\%$ with normal visible mass making a smaller percentage of about a $20\%$ contribution. Thus the important conclusion is that {\it dark energy\/} value within the universe is a direct consequence of the induced mass from the $\Delta$ field which itself is largely {\it dark mass\/}.  Briefly, dark energy within the universe is numerically very close in value to the {\it vacuum polarised\/} dark mass and if the $\Gamma$ field is also classified as dark the closeness becomes coincidence.
From the preceding discussion and equation (\ref{B53}) it should not be inferred that dark mass is a primary source of dark energy. I think the reverse is nearer to the truth and equation (\ref{B53}) is the direct result of a mechanical equilibrium between pressure equivalent induced density from the  $CMB$ and the sum of the pressure induced densities from the $\Delta$ and $\Lambda$ field at the boundary and within the universe. Thus this mechanical equilibrium effectively transfers the dark energy pressure from outside the universe to its boundary and hence by homogeneity to inside the universe. The $PEID$ concept will be explained in the next section on pressure equivalent induced densities.
\section{Pressure Equivalent Induced  Density, PEID}
\setcounter{equation}{0}
\label{sec-peid}
It turns out to be very useful to introduce the concept of {\it Pressure Equivalent Induced Density, {\it PEID\/}}, in relations to the equations of state associated with specific subsystems of the total system. For example, suppose one subsystem is called the $\Delta$ system with the equation of state,
\begin{eqnarray}
 P_\Delta (t) &=& c^2\rho_\Delta\omega_\Delta (t)\label{C0}\\
\Delta (t) &=& \rho_\Delta (t) \omega_\Delta (t) \label{C1}\\
 &=&P_\Delta (t)/c^2,\label{C2}
\end{eqnarray}
then I take the definition for the $PEID,\ \Delta (t),$  to be given by equation (\ref{C1}). Thus $\Delta (t)$ has the same dimensions as density because in common with all the omegas, $\omega_\Delta (t)$, is dimensionless and it is derived from $ \rho_\Delta (t)$ through the multiplicative action of the inducing function, $\omega_\Delta (t)$. From (\ref{C2}) it is clearly essentially a pressure with the dimensions of density.  It represents this pressure in the form of the {\it mass density\/}, $\Delta (t)$. I am not aware that the $PEID$ slant on equations of state has any important part elsewhere in physics but it seems that it does play an essential role in cosmology in relation to the understanding of dark energy and its connection to other key densities. This is clear from inspection of equation (\ref{B0}) again with and without the $G$ weightings,
\begin{eqnarray}
G\rho_\Lambda &=& G_{-} \Gamma_B(t)   + G_{+} \Delta _B(t)\label{C3}\\
\rho_\Lambda &=& \Delta _B(t)- \Gamma_B(t).\label{C4}
\end{eqnarray}
Thus from equation (\ref{C4}) the source of dark energy density within the universe is just the difference of the $PEID$s for the $\Delta$ and $\Gamma$ fields which together constitute all the conserved mass of the universe. Thus the mystery of the origin of the dark energy density, $ \rho_\Lambda = \Lambda c^2/(8\pi G)$ in Einstein's form or in my revised form $\rho^\dagger_\Lambda=2\rho_\Lambda $, within the universe is completely resolved by this theory. Possibly this is the reason that dark energy is not visible. It could be because $pressures$ are not usually visible and the {\it pressure status\/} of the dark energy density is its dominant characteristic. However, it seems to me that dark energy with approximately an equivalent density of $5$ hydrogen atoms per cubic meter would not be visible anyway. The formula (\ref{C4}) can also be used to show a simple relation between {\it dark mass\/} and  {\it dark energy\/} but before discussing that aspect it is useful to consider in the next paragraph the way this theory structure has developed and can continue developing.
In the first two papers, {\it A and B\/} of the four {\it A,B,C,D\/}, I found the dust universe model from scratch by just integrating the Friedman equations. The result subsequently turned out to be a reincarnation of the first model introduced by {\it Lema\/}$\hat\imath${\it tre\/} \cite{47:lem}  but with substantially different interpretations and additional details. The version of the model in {\it A and B\/}, like most cosmological models, involved the assumption that the mass density of the universe only depended on time and so was space-wise homogeneous. However, the structure unearthed in that version of the model was completely adequate to describe cosmological expansion and its change from deceleration to acceleration at some time $t_c$ in the past and various other new understandings of the cosmological process, all in complete agreement with up to date measurement. Thus this basic structure did not depend on differentiating the mass density into separate components to represent various contributory fields such as the electromagnetic or heavy particle contributions. The dark energy contribution was involved in that version of the theory but not included as part of the conserved mass of the universe, it was rather treated as a permanent constant density resident of the hyperspace into which the universe expands. I shall here denote that model by $U_0=U_\Lambda(DM)$, meaning that it can be assumed to {\it only\/} contain an {\it energy conserved over all time\/} quantity of {\it dark mass}, $M_U$, while, as we have seen, it swims in and is permeated with the dark energy content of an enveloping 3D-hyperspace. The conserved mass density, $\rho (t) \sim \Omega_M (t)$, in this model {\it must\/} represent all the dark mass, if we assume that none of this dark mass has converted into visible mass and further because it satisfies the equation (\ref{B27}) which has to add up to unity to ensure that fact. Thus the model $ U_\Lambda(DM)$, can be conceived as not containing any visible hadronic matter, which as we know can only be present in a very small proportion anyway and it would also likely be none uniformly distributed. It follows that the model $U_0=U_\Lambda(DM)$ can be regarded as a very bland, over all time, approximation to the actual universe and which can be built up in stages to represent the universe with increasing accuracy. I emphasise the usual cosmological basic assumption that the model's density function is space-wise homogeneous means that if the model contains any dark mass within its boundaries then it contains {\it only uniform\/} dark mass and together with the uniformly distributed dark energy background. The next stage in the build up process in which the cosmic microwave back ground was added was published in {\it C\/} and will be denoted by $U_1=U_\Lambda (DM =\Delta (t)\cup\Gamma (t) )$. This means that the fixed amount of dark mass in the first version is now able to transform into time dependent components $\Delta (t)$ for one part and $\Gamma (t)$ for the complementary part, the $CMB$, with the same total mass quantity as the original dark mass. The next stage of complexity is the introduction of the possibility that part of the $\Delta$ mass, $M_U$ can transform into visible mass, often called hadronic mass. This universe can be represented by $U_2=U_\Lambda (DM =(\Delta (t)=\Delta_D (t) \cup \Delta_V (t))\cup\Gamma (t) )$ with now the quantity of $\Delta$ mass being shared between the dark and visible versions as denoted by the $D$ and $V$ subscripts. Clearly the increasing complexity procedure can continue to produce universes with lower homogeneity described by $U_3$ and so on. Let us now return to discussing the relation between dark mass and dark energy.
\section{Dark Mass, Dark Energy Ratio}
\setcounter{equation}{0}
\label{sec-dmde}
Consider firstly the basic universe type Friedman dust universe, $U_0$. The model in this basic case is an excellent representation of the modern astronomical measurements. However the basic density function is assumed to be rigorously homogeneous and contains only conserved with time dark mass and the hyperspace permeating dark energy. The density functions for the dark mass, dark energy  and the ratio, $r_{\Lambda,DM} (t)$, of dark energy to dark mass as functions of time are respectively represented by
\begin{eqnarray}
\rho (t) &=& (3/(8\pi G))(c/R_\Lambda)^2\sinh^{-2}(3 c t/(2R_\Lambda))\label{C5}\\
\rho^\dagger_\Lambda &=& (3/(4\pi G))(c/R_\Lambda)^2\label{Cc6}\\
r_{\Lambda,DM} (t) &=&\rho^\dagger_\Lambda /\rho (t) = 2 \sinh^{2}(3 c t/(2R_\Lambda)) \label{C7}\\ 
r_{\Lambda,DM} (\pm t_c) &=& 2 \sinh^{2}(\pm 3 c t_c/(2R_\Lambda))= 1.\label{C9}
\end{eqnarray}
Equation (\ref{C7}) is a general result but in the case of  a $U_0$ universe it can be expressed differently by using equation (\ref{C4}) with the $\Gamma$ term taken zero as 
\begin{eqnarray}
\rho_\Lambda &=& \Delta _{B,0}(t)\label{C10}\\
&=& \rho(t)\omega_{\Delta ,0} (t),\label{Cc11}
\end{eqnarray}
the zero subscripts having been added to differentiate the functions concerned from those in the $U_1$ version.
From paper $C$, we know that
\begin{eqnarray}
\omega_\Delta (t) &=& \left(\frac{M_\Gamma }{3 M_U} +\frac{3 (c/R_\Lambda)^2 \rho^{-1}(t)}{8 \pi G} \right)/(1 -M_\Gamma/M_U).\label{c12}
\end{eqnarray}
Thus the zero $\Gamma$ version for $U_0$ is given by
\begin{eqnarray}
\omega_{\Delta,0} (t) &=& \left(\frac{3 (c/R_\Lambda)^2 \rho^{-1}(t)}{8 \pi G} \right).\label{C13}
\end{eqnarray}
Substituting this into equation (\ref{C11}) confirms the validity of (\ref{C11}).
Thus the rather trivial equation (\ref{C11}) gives the {\it all\/} time dependent relation between dark energy and dark mass for the nontrivial model $U_0$. However, trivial or not, the dark energy and dark mass densities are {\it strongly\/} numerically related through the function $\omega_{\Delta,0} (t)$ and this applies for all time, ($-\infty < t < + \infty$).
Let us now consider the ratio, $r_{\Lambda,DM} (t)$, of dark energy to dark mass in the case of a universe in which the homogeneity has been broken by the addition of the cosmic microwave background, replacing some by the $CMB$.
From (\ref{C7}), we have generally,
\begin{eqnarray}
r_{\Lambda,DM} (t) &=&\rho^\dagger_\Lambda /\rho (t) = 2 \sinh^{2}(3 c t/(2R_\Lambda)). \label{C14}
\end{eqnarray}
However, with the addition of the $\Gamma$ field
\begin{eqnarray}
\rho (t) = \rho_\Delta (t) + \rho_\Gamma (t)\label{C15}
\end{eqnarray}
so that the dark energy dark mass ratio of $U_0$ at (\ref{C14}) becomes in $U_1$
\begin{eqnarray}
r_{\Lambda,DM,1} (t) &=& \frac{\rho^\dagger_\Lambda }{\rho_\Delta (t) + \rho_\Gamma (t)} = 2 \sinh^{2}(3 c t/(2R_\Lambda)). \label{C16}  
\end{eqnarray}
The denominator of the ratio remains unchanged as also does the second equality because the numerical values are unchanged. It might be thought that the left and right sides of the first equality do not now agree because only the $\Delta$ part contains dark mass, that which is left from the $U_0$ universe case after some has converted to $CMB$. Numerically there is no problem as the quantity of dark mass is presumably shared between the $\Delta$ and $\Gamma$ fields. However, the terminology might be questioned. {\it Arguably\/}, the $CMB$ is composed of photons which are not visible and therefore the $CMB$ can be classified as {\it dark mass\/} equivalent material. Of course photons convey information about {\it other visible\/} materials to the eye but photons themselves are not {\it seen\/} in the usual meaning of the word. I have added the extra subscript $1$ in the $U_1$ ratio so that no confusion can arise if the case I have just made is not accepted.
The dark energy dark mass ratio in either form above represents a {\it fundamental\/} time conditioned relation between dark mass and dark energy.
This result and the formula (\ref{C4}) both of which hold inside and on the boundary of the universe show how totally interdependent are the two {\it dark\/} facets. The ratio $r_{\Lambda,DM} (t)$ is of great generality and could play an important part in helping to understand cosmological {\it voids\/}, a recent astronomical discovery. This ratio has come out of general relativity but it can be shown that it is independent of general relativity and its existence only depends on some simple assumptions added to Newtonian gravitational theory. The very basic and major significance of this ratio will be discussed and demonstrated in the next section by showing that it is directly derivable from Newtonian gravitational theory. It will be indicated how this implies a context for its significance within smaller regions of space within the universe's boundary. 
\section{Newtonian Dark Mass and Dark Energy}
\setcounter{equation}{0}
\label{sec-ndmde}
Consider an infinitely extended 3-dimensional Euclidean space such as that in which Newtonian gravity is usually considered to act between objects having the physical characteristic called mass. I shall make the usual assumption that Newtonian gravity acts between enclosed regions of space of spherical shape that enclose a {\it uniform\/} density distribution of mass that can change with time but retaining an overall {\it fixed\/} quantity with respect to time of the usual positive gravitational mass within it boundary, an amount $M$, say. Usually there will be some moving gravitational centroid at which the gravitation force between objects will be thought to be acting. I also only use configurations in which this centroid is the centre of a sphere. The difference from Newtonian theory that I am about to introduce is the assumption that this Euclidean space is filled uniformly throughout all its extent by a positively mass density field of negatively characterised gravitational material such as the dark energy found to exist in the cosmos. This negative gravity material will be denoted by the constant density, $\rho^\dagger _\Lambda =c^2\Lambda/(4\pi G)$ just as in my double version of the Einstein theory quantity, $\rho_\Lambda =c^2\Lambda/(8\pi G)$.
Consider now a spherical region of this space of radius $r$ about the origin of this space as centre. Suppose this sphere contains a total amount of {\it dark mass\/}, $M$, with its positive gravitation characteristic, $G$. The sphere will also contain an amount of negative gravity, $-G$, dark energy given by
\begin{eqnarray}
M_\Lambda &=& \rho^\dagger_\Lambda V (t)\label{C17}\\
V(t) &=&4 \pi r^3(t)/3.\label{C18}
\end{eqnarray}
Thus the total gravitational acceleration caused by the sphere's contents at its surface will be given by the Newtonian gravitational formula,
\begin{eqnarray}
\ddot r (t) &=& M^\dagger _\Lambda G/r^2(t) - MG/r^2(t) \label{C19}\\
&=& 4\pi r^3 \rho^\dagger _\Lambda G/(3 r^2) - C/(2r^2) \label{C20}\\
&=& 4\pi r \rho^\dagger _\Lambda G/3  - C/(2r^2) \label{C21}\\
&=& rc^2\Lambda /3 - C/(2r^2). \label{C22}
\end{eqnarray}
If we multiply equation (\ref{C21}) through by $\dot r$, we obtain
\begin{eqnarray}
\ddot r \dot r &=& 4\pi r\dot r \rho^\dagger _\Lambda G/3  - C\dot r/(2r^2)\label{C23}\\
\frac{d}{dt}{\dot r}^2/2&=&\frac{d }{dt} r^2\Lambda  c^2/6 + C \frac{d}{dt} r^{-1}/2 \label{C24}\\
{\dot r}^2&=& (rc)^2\Lambda /3 + C r^{-1}  \label{C25}\\
C &=& 2MG .\label{C26}
\end{eqnarray}
The constant of integration that could occur in integrating (\ref{C24}) can be taken to be zero under the conditions that $\dot r(t)$ is taken to be infinite with $r(t)=0$ at $t=0$. Thus the spherical region expands with high speed from the origin, $r=0$ at time $t=0$. 
The solution to equation (\ref{C25}) was obtained in paper $A$ in the form
\begin{eqnarray}
r(t) &=& b\sinh^{2/3}(3 ct/(2R_\Lambda))\label{C27}\\
R_\Lambda &=& (3/\Lambda)^{1/2}\label{C28}\\
b &=& (R_\Lambda /c)^{2/3} C^{1/3}  \label{C28.1}\\
C &=& 2 M G\label{C28.2}
\end{eqnarray}
where $M$ here is any {\it dark mass\/} value.
It follows that the {\it dark\/} mass density of the spherical region containing total dark mass, $M$, is as in (\ref{C5}) given by
\begin{eqnarray}
\rho (t) &=& M/(4\pi r^3(t)/3)= M \sinh^{-2}(3 ct/(2R_\Lambda))/b^3\label{C29}\\
&=& (3/(8\pi G))(c/R_\Lambda)^2\sinh^{-2}(3 c t/(2R_\Lambda)).\label{30}
\end{eqnarray}
Thus the ratio of dark energy {\it mass\/} density to dark mass density within this region over time is
\begin{eqnarray}
r_{\Lambda,DM} (t) = \rho^\dagger_\Lambda /\rho (t) = 2 \sinh^{2}(3 c t/(2R_\Lambda)) \label{C31}
\end{eqnarray}
which again is the same as (\ref{C7}).
The formula for the ratio of dark energy to dark mass, $r_{\Lambda,DM} (t)$, depends only on the dark mass density through $t$ and $R_\Lambda$. The time variable origin $t=0$ depends only on where the sphere expansion is assumed to have started from with radius zero, an arbitrarily chosen space origin $r(t)=0$ at time $t=0$, in Euclidean three space. Thus it seems that this is a {\it fundamental\/} formula governing a time evolutionary process relating dark energy and dark mass. The consequence of this situation is that we can visualise, quite independently of relativity, such mixed mass region expansions.
They can take place over time from anywhere in astro-space and {\it apparently\/} originate from a point quantity of {\it dark mass\/}, $M$, with infinite density. Further, the formula is time reversible so that it suggests that spherical contractions of spherical dark mass regions can also be visualised as a possible cosmological sequence of events resulting in the appearance of a point dark mass, M, with infinite density locally. As such an expansion proceeds the spherical region picks up dark energy mass from the enveloping Newtonian space, the expansion continuing with the expanding region having then a mixture of the two gravitational types of mass, $\pm G$. An important event in the history of such an expansion is when there are equal quantities of the two mass types within the sphere. At this event occurring, the sphere will be gravitationally neutral. The sphere will at that time exert no gravitational force on material outside its boundary, it will be gravitationally isolated from any material exterior to itself. If we denote the time when the sphere is so isolated by $t_c$ this time can be found from the formula of dark mass and dark energy mass equivalent equality, either equation (\ref{C32}) or equation (\ref{C33})
\begin{eqnarray}
r_{\Lambda,DM} (t_c) &=& \rho^\dagger_\Lambda /\rho (t_c) = 1\label{C32}\\
\rho^\dagger_\Lambda &=&\rho (t_c) \label{C33}\\
\sinh ^2 (3ct_c/(2 R_\Lambda)) &=& 1/2 \label{C34}\\
\Rightarrow t_c &=& \pm (2R_\Lambda /(3c))\sinh ^{-1} (1/2^{1/2})\label{C35}
\end{eqnarray}
and, curiously, the times $\pm t_c$  do not depend on the amount of dark mass within the expanding sphere but only depends on the value of the cosmological constant, $\Lambda$. It follows that the time $t_c$ has exactly the same value as the relativistic epoch time when the universe changes from deceleration to acceleration. The time $t_c$ is a {\it fundamental\/} universal time interval in the cosmological context. It is important to note that, as the process is time reversal invariant, the contraction sequence, in negative time, with mass $M$ can be immediately followed by an expansion sequence with the same mass $M$, in positive time, so that conservation of mass is assured and mass is neither created from nothing nor is it destroyed at the singular event when $t=0$. The non dependence of the process on the amount of dark mass within the boundary of the contracting or expanding sphere of dark mass has a surprising explanation. The process conforms exactly to the {\it principle of equivalence\/}. Just as the acceleration of a falling mass in a gravitational field does not depend on the value of the falling mass so the acceleration $ \ddot r_{\Lambda,DM} (t)$ of the collapsing sphere process does not depend on its mass. The collapsing sphere in its own gravitational field conforms exactly too and is a manifestation of the principle of equivalence. It can occur locally and is a basic part of the description of the whole universe motion with epoch time. Recognition of this fundamental process in relation to other physical processes in cosmology will be discussed in the final section. 
\section{Appendix 1 Conclusions}
\setcounter{equation}{0}
\label{sec-app1con}
The cosmological model introduced in references $A$, $B$, $C$ and applied to the finding of solutions to the {\it cosmological constant\/} problem in $D$ has here been applied to unravelling the {\it dark mass\/} problem. Here it has been shown that a fundamental time moving relation holds between {\it dark energy\/} and {\it dark mass\/}. This relation was first shown to hold at the scale of the whole universe by using the Friedman equations from Einstein's general relativity and involving his positively valued cosmological constant $\Lambda$.
Here it has been shown that the same relation can be derived from Newtonian gravitation theory with only the addition of a constant and universally distributed density of dark energy, $\rho^\dagger_\Lambda = 2\rho_\Lambda$, twice the Einstein value $\rho_\Lambda $, in Newtonian space and only subject to Newtonian gravity theory. This result implies that the formula relating dark mass and dark energy is independent of general relativity and the way it is derived also show that it can have applications at a much smaller scale than that of the entire universe. It can describe local space and time  {\it small scale\/} movements of dark mass in relation to dark energy. Thus I suggest the formula could play an important role in explaining the way that dark mass, if taken to be primary positive gravity, $+|G|$, mass, can condense, precipitate or clump to become galaxies or just empty voids  \cite{59:rud} in the cosmological fabric. As we have seen, there are five main events in the time sequences evolution of this dark energy dark mass process, $E_0,E_{\pm 1},E_{\pm\infty}$, say.  They involve $E_0$ when some definite random quantity of dark mass $M$ is located at some definite point in three space at some definite time labelled as $t=0$ for the process. At that time the dark mass is by itself because a point cannot contain any of the uniform and finite constant density of dark energy mass. Thus in space around the point mass it will own a Newtonian gravitational potential field $-MG/r$. At both the events $E_{\pm 1}$ at times $\pm t_c$ because of the time reversal invariance the contracting or expanding sphere will contain equal quantities of the dark mass and dark energy so that the sphere will be gravitationally neutral. It will thus be isolated gravitationally and so not own any gravitational potential. However the total mass density within the spheres boundaries will be $\rho (t_c) +\rho^\dagger _\Lambda$, a numerically very small value $\approx 9$ proton masses per cubic meter. I think that such a sphere being gravitationally isolated and of such low density could qualify for the title {\it cosmological void\/}. At the events $E_{\pm\infty}$, the sphere will own a gravitation potential at points within its surface involving both the dark energy and dark mass within concentric spheres of radius $r < \infty$ but dominated by the repulsive dark mass for relatively large values of $r$. The contraction phase between $E_{-\infty}$ and $E_0$ might represent a moving platform for an original dark mass concentration to convert from pure dark mass to becoming dark mass contaminated with visible mass while its volume descends to occupying some relatively small region containing a group of visible galaxies or, a single galaxy or even a single particle. In other words, the descending spherical volume could represent a time dependant packaging process for cosmological clumping. A final remark about the relation of this theory structure to {\it aether theory\/} is appropriate. It is dark energy rather than dark mass that seems to play a role much like the all pervading aether which has been used to give a physical explanation for electromagnetic wave motion in so called  {\it empty\/} space. The dark energy density is certainly an all-pervading effect in this cosmological theory as has been shown in this article and as it is also perceived in the present day arena of astronomical observations.  It seems to be an everywhere present background reference level against which many astrophysical and quantum problems can be understood and measured.  The dark mass or positive gravity element appears to represent a measure of a soliton like wave effect either universally or locally of a boundary motion at an interface between dark mass and dark energy described by the inverse of the ratio, $r_{\Lambda,DM} (t)$.

\section{Appendix 2}
\setcounter{equation}{0}
\label{sec-app2} 
\vskip0.6cm 
\centerline{\Large {\bf Expanding Boundary Pressure Process}}
\centerline{\Large {\bf All Pervading Dark Energy Aether}}
\centerline{\Large {\bf in a Friedman Dust Universe}}
\centerline{\Large {\bf with Einstein's Lambda}}
\vskip 0.5cm
\centerline{\bf Abstract}
\vskip 0.7cm
In this appendix a more detailed study of the role of dark energy mass as a conserved with time substance  that permeates the expanding universe. It shown that if dark energy is to be conserved over all time it has to satisfy the cosmological vacuum polarisation equation over the pre-singularity range of contraction and the post-singularity range of expansion in order for it to remain in a self mechanical equilibrium inside and outside the boundary of the expanding universe and so be able to be always and everywhere permeable to the positive gravity {\it dark mass\/} and visible mass material within the universe. 
\vspace{0.4cm}
\section{Effect of Boundary Pressures}
\setcounter{equation}{0}
\label{sec-intro2}
In the paper $D$, \cite{58:gil}, it was shown that the quantum vacuum polarisation idea can be seen to play a central role in the Friedman dust universe model introduced by the author. An essential part of that role involves the relations between three pressures at the boundary of the expanding universe. In particular of fundamental importance is a relation between pressure from the $CMB$, $P_\Gamma$, pressure from all the rest of the universe which is not $CMB$ and not dark energy, $P_\Delta$, and pressure from dark energy itself, $P_\Lambda$. This relation takes the form
\begin{eqnarray}
P_\Gamma  &=& P_\Delta + P_\Lambda\label{E0}\\
P_\Lambda &=& c^2\rho_\Lambda\omega_\Lambda =-c^2\rho_\Lambda \label{E1}\\
\omega_\Lambda &=& -1.\label{E2}
\end{eqnarray}
Equation (\ref{E1}) together with equation (\ref{E2})  is the well known specification implying negative pressure, $P_\Lambda$, from the dark energy density in the Einstein form $\rho_\Lambda = \Lambda c^2/(8\pi G)$. In earlier work, I have referred to the equation (\ref{E0}) as representing a mechanical equilibrium between the $\Gamma$ field and the $\Delta$ and $\Lambda$ fields combined. I now think that designation while remaining {\it formally\/} correct should be presented with a changed interpretation because of the negative pressure associated with the dark energy field, $\Lambda$. In the usual specification of a mechanical equilibrium two pressures $P_1 =P_2$ are said to be equal where there is no complication of possible negative parts for either of them. Pressures on either side of a boundary between non-miscible liquids for example are said to be in mechanical equilibrium if the boundary is not accelerating. In such a case, although the pressures act at the boundary in opposite directions they are both taken as positive. Mechanical equilibrium in thermodynamics is a very contentious area of research so that my explanation in this context is very minimal. As a result of these complications it is desirable to express equation (\ref{E0}) in the alternative form using a modulus sign, $|\ |$. The term {\it boundary\/} of the universe at time $t$ refers to a conceptual sphere of radius given by the function $r(t)$ defined earlier (A).  
\begin{eqnarray}
P_\Gamma  &=& P_\Delta -|P_\Lambda |\label{E3}\\
P_\Delta &=& P_\Gamma + |P_\Lambda |.\label{E4}
\end{eqnarray}
In this form, all the pressures are expressed as positive quantities and the mechanical equilibrium between these three field can now be more safely reinterpreted as a mechanical equilibrium between the $\Delta$ field and the $\Lambda$ and $\Gamma$ fields combined. This version of the equilibrium condition at the boundary of the expanding universe makes good sense physically for at least two reasons.  The first reason is that dark energy material exists on both sides of the expanding boundary of the universe so the $\pm|P_\Lambda|$ versions refer to the pressure direction on the boundary from dark energy material on one side or the other, whilst the gravitational pressure, $P_\Delta $, is directed towards the material within the universe and so, on the boundary, is only effectively equivalent to a positive pressure towards the centre of expansion. This last property is a well-known result originating from Newtonian gravitational theory.  Taking the total pressure, $P$, in the Friedman equations as positive is a rather anomalous convention which has caused much confusion which I attempt here to unravel. The second reason is that the form (\ref{E3}) rearranged as in (\ref{E5})  
\begin{eqnarray}
P_\Lambda &=& -|P_\Lambda |=  P_\Gamma -P_\Delta  \label{E5} \\
P(t)&=&+P_G + P_\Lambda = P_\Delta (t) -P_\Gamma (t) + P_\Lambda\equiv0\  \forall\  t.\label{E5.1}
\end{eqnarray}
clearly expresses the physics of the equilibrium condition. It is that the negative outward pressure $P_\Lambda $ that would be exerted on the boundary from the dark energy inside the universe is equal to the difference of the outward $CMB$ pressure, $P_\Gamma $, less the inward pressure $P_\Delta$ exerted on the universe boundary from within by the positive $G$ or normal gravitating material within the universe. The equations (\ref{E5}) and (\ref{E1}) firmly identify both the pressure $P_\Lambda$  and the mass density $\rho_\Lambda$ of the dark energy as coming from the quantities $ P_\Gamma $ and $ P_\Delta $ both defined  with meanings within the universe. Elsewhere, I have expressed the equation (\ref{E5}) using the $PEID$ form which explains it in terms of the mass densities, $\Gamma (t),\ \Delta (t)$ rather than the equivalent pressures,
\begin{eqnarray}
\rho_\Lambda &=& \Delta (t) -\Gamma (t). \label{E6}\\
G\rho_\Lambda &=& G_+\Delta (t) +G_-\Gamma (t) \label{E7}\\
G_+&=&+G\label{E8}\\
G_-&=&-G.\label{E9}
\end{eqnarray}
The equation (\ref{E5.1}) uses (\ref{E5}) to bring us back to the total pressure $P(t)$ which as indicated is identically {\it zero\/} for all $t$ and so indicates that the whole history of the universe in this model is that of a {\it dust\/} universe.  Clearly then the total pressure $P(t)$ cannot be responsible for the acceleration. This conclusion agrees with what was noted earlier that the acceleration is accurately determined by a generalisation of the Newtonian gravitation theory only involving adding to the inverses square law a linear law term involving Einstein's cosmological constant $\Lambda$. The realisation that the various pressures that we have been discussing earlier do not determine the dynamical behaviour of the system generates the question, {\it what is this complicated pressure structure all about?\/}

The immediate answer to the question at the end of the last paragraph is that the pressure structure that derives from the equilibrium relation between the three pressures, $P_\Lambda$, $P_\Delta$ and $P_\Gamma$  generates the important relation, (\ref{E6}) or (\ref{E7}). This relation shows that {\it within\/} the spherical volume of the universe the constant valued dark energy density, $\rho_\Lambda$, is determined by the {\it dark mass\/} dominated quantity pair $\Delta (t)$ and $\Gamma (t)$. I emphasise {\it within\/} because both these quantities are part of the constant space and time conserved energy of the universe, $M_U$. However, the space-time constant {\it dark energy\/} mass density $\rho_\Lambda$, by initial assumption, exists everywhere in the universe's enveloping space with the same definite numerical value outside as inside. Thus the following subsidiary question presents itself: If the dark energy density inside the universe is given by (\ref{E6}) in terms of the internal constituents, $\Delta$ and $\Gamma$, how is it that outside the universe involving regions which will not have been reached by the internal constituents of the expanding universe, the dark energy density $\rho _\Lambda$ exist in its own right by assumption, with the same constant value as inside  and apparently not generated by any internal influence?  The unique character of this model does allow a satisfactory answer to this question which depends on the model's strict conformance to the principle of conservation of energy in contrast with the {\it standard  big bang model\/}. This model involves two basic {\it positive\/} types of mass defined by their gravitational character, which is determined by whether the mass appears in the theory multiplied with $G_+ =+G$ or $G_-=-G$,  where the gravitational constant $G$ is always constant, $G>0$. Dark mass and normal mass belongs to the $G_+$ category and dark energy belongs to the $G_-$ category. An important way in which this model differs from the big bang type universe is that the beginning of time in this theory, rather than occurring at time $t=0$, occurs at time $t=-\infty$ and the end of time occurs at $t=+\infty$.  This can be interpreted as this universe lasts {\it for ever\/}.  The reader may prefer to regard this infinite time scale as just one out of a possible infinite number of infinite contiguous periodic time scales and so reinforcing the lasting for ever concept. This last extension can be usefully incorporated in the theory, see paper (C). Let us now consider the situation at and after the start of time taken as $t=-\infty +t_\epsilon $, $ t_\epsilon \approx +|0|$ at this stage the radius of the universe is infinitely large, $r(-\infty +t_\epsilon)$,  and will decrease with  advancing time. In other words, near the beginning of time the universe is a sphere occupying almost the whole of hyperspace and so the internal generating dark energy process (\ref{E6}) is operative almost everywhere in hyper space. It follows that the universe is full of the dark energy density $\rho_\Lambda$, though this density is itself small it adds up in total over the whole universe volume to a {\it very large amount\/} of dark energy mass. The universe also contains a much smaller density value, $ \rho(-\infty +t_\epsilon)$, of conserved positive gravitational mass, $M_U$,  so that $\rho(-\infty +t_\epsilon)V_U(-\infty +t_\epsilon) =M_U$. This conserved mass contains the mass of the universe that we see. The basic assumption in this model is that dark energy density, $\rho _\Lambda (t)$, exists everywhere and at all time so that if the radius of the universe at $t=-\infty$ is infinite and  if the space is flat Euclidean then the universe has no outside and takes in all the hyper-universe so that all the dark energy is enclosed. After the small time elapse, $t_\epsilon $, the universe will have acquired a small outside volume and a slightly smaller inside volume than it had initially. It follows that in principle there are two types of simple likely possibilities. Firstly, the contracting universe leaves no dark energy density outside as it evolves in time and keeps the original value inside at the same value as given by formula (\ref{E6}).  Secondly, as it evolves in time it leaves outside sufficient dark energy density to keep to the uniform constant density condition everywhere and so keeping the dark energy density inside at the same value given by formula (\ref{E6}) as outside.  The first option means that the dark energy within the universe would decrease with decreasing volume consequently losing density to no recognisable sink and so implying dark energy is not conserved. This would also violate the assumption that dark energy density is constant everywhere and at all time. Thus we are left with only the second possibility and consequently the actual scenario has to be that as the universe contracts the internal pressure process described by the internal $\Delta$ and $\Gamma$ fields precipitates the right amount of dark energy material outside in the space produce by the contracting universe. This process will continue for all time and, in particular, past the singularity at $t=0$ when the volume is zero. Thus after the singularity, when the universe is in an expanding mode, it will encounter the pre-singularity dark energy density outside its boundaries precipitated in its contracting mode. Thus the main role of formula (\ref{E7}) is to keep the conserved and bounded  dark mass within the universe freely permeable to or non interacting with the dark energy in which it swims by maintaining the self mechanical equilibrium of the dark mass density in the form,
\begin{eqnarray}
P_{\Lambda ,in} =P_{\Lambda, out}  \label{E10}\\
\rho_{\Lambda ,in} =\rho_{\Lambda, out}, \label{E11}
\end{eqnarray}
where $P_{\Lambda ,in}$ is the dark energy pressure just inside the boundary and $P_{\Lambda , out}$ is the dark energy pressure just outside the boundary of the universe. Equation (\ref{E11}) gives the same condition in terms of densities.
\section{Appendix 2, Conclusions}
All this suggests that the dark energy density, although pressure identified, has also to be taken seriously as a genuine mass density. It also shows that the pre-singularity negative time phase is a necessary adjunct to making sense of this theory. The conclusion associated with this section is that the formula (\ref{E7}) together with the full time history of this model assures that dark energy and dark mass are both conserved over all time. The above argument is not meant to be mathematically rigorous but rather a plausibility construction. No doubt the reader can think of various improvements
\section{Appendix 3}.

\vskip 0.2cm

\large   
\centerline{\Large {\bf A Quantum Theory Friendly Cosmology}}
\centerline{\Large {\bf Exact Gravitational Waves}}
\centerline{\Large {\bf based on a Friedman Dust Universe}}
\centerline{\Large {\bf with Einstein's Lambda}}
\vskip0.5cm
\large \centerline{\bf Appendix 3 Abstract}
\vskip 0.2cm
In this paper, it is shown that this cosmological model can be expressed in a form which is quantum theory friendly. That is to say, besides not have a cosmological constant problem and also not having a coincidence problem, aspects dealt with in earlier papers and continued in the first part of this paper, it is shown that the dust universe can be expressed in a form having a close resemblance to the Schr\"odinger equation formalism. This resemblance cannot be seen as an identity of the two systems because the Schr\"odinger equation is linear and the Friedman equations are non-linear. This aspect is discussed in detail and a precise relation is shown to exist and is demonstrated to hold between cosmology theory structure  and the quantum theory linear superposition of eigen-states. This relation describes  cosmology's non-linearity relative to Schr\"odinger linearity and is called, bilinear superposition. However, in spite of not achieving an identity of structure between cosmology and quantum theory, sufficient equivalence can be shown to exist via a comparison of quantum wave motion as described by the Schr\"odinger equation and gravitational wave motion as described by the Friedman dust universe to suggest that a quantum theory of cosmology and gravity is likely to be possible via this route. An exact {\it non-linear\/} Schr\"odinger  equation description for the model is obtained.
In this paper's appendix, it is shown that this Schr\"odinger equation has an infinite multiplicity of space variable solutions that can be used to remove the usual restriction of cosmology theory to uniform space variation with dependence on epoch time only.    
In the paper $D$, \cite{58:gil}, it was shown that the quantum vacuum polarisation idea can be seen to play a central role in the Friedman dust universe model introduced by the author. In the paper, \cite{64:gil}, it was shown that the Friedman equation structure can be converted into a {\it non-linear\/} Schr\"odinger equation structure. Here, this aspect is further developed by supplementing the solutions to this time only equation with a  dependence on a three dimensional  space position vector, {\bf r}, so that the  equation remains consistent with its cosmological origin. This step then enables finding cosmological models that are not restricted to having a mass density that is certainly time dependent but otherwise remains constant over all {\it three dimensional\/} position space at every definite time.
 
Altogether, the objective has been to produce an alternative to the standard model which contains less paradoxical structure than does the standard model and which at the same time is hopefully adaptable to being quantized in some sense or other. The question of in {\it what sense is the model to be quantized\/}, I see to be an open question which may or may not have a unique answer and to this issue discussion will here be devoted. My strategy is to mould the cosmological model of the Friedman dust universe into a form that has a structure as near as possible to the structure of Schr\"odinger quantum theory by emphasising a wave motion aspect of the dust universe Friedman model. This will be explained in detail in Section {\bf 5} which is devoted to consideration of the well known essentials of Schr\"odinger quantum theory that need somehow to be present in the cosmological model. 
The basic version of this dust universe model is described by a sphere in three dimensional Euclidean space with a changing with epoch time, t, radius magnitude, r(t),
\begin{eqnarray}
r(t) &=&  b\sinh^{2/3}(\pm 3ct/(2R_\Lambda))\label{t1}\\
b &=&  (R_\Lambda /c)^{2/3} C^{1/3}\label{t2}\\
R_\Lambda &=& (3/\Lambda )^{1/2}\label{t3}\\ 
C &=& 2M_UG.\label{t4}
\end{eqnarray}
$R_\Lambda$ is often called the de Sitter radius and $C$ is Rindler's constant formed from twice the product of the conserved constant mass, $M_U$, within the total changing volume of the universe and the Newtonian gravitational constant, $G$. $\Lambda$ is Einstein's cosmological constant. $r(t)$ is rigorously a solution to the Friedman equations and consequently also rigorously a solution to Einstein's field equation's which holds for epoch time from $t=-\infty$ to $+\infty$. The $\pm$ can usually be omitted provided the cube root of the $\sinh$ function is assumed taken after squaring so that radius, $r(t)$, is time symmetry invariant, $r(t)=r(-t)$ and no complex roots are involved. The formulae (\ref{t1})$\rightarrow$ (\ref{t4}) contain all the basic mathematical-theoretical information about the Friedman dust universe model involved in this research program. For example, the Hubble {\it function\/}, $H(t)$, the conserved mass density, $\rho(t)$, and Einstein's dark energy density, $\rho_\Lambda$,  are given by
\begin{eqnarray}
H(t) &=& \dot r(t)/r(t) = (c/R_\Lambda ) \coth(3ct/(2R_\Lambda))\label{t1.1}\\
\rho (t) &=& 3 M_U/(4 \pi r^3(t))= \rho_\Lambda \sinh^{-2}(3ct/(2R_\Lambda))\label{t2.2}\\
\rho_\Lambda &=& \Lambda c^2/(8\pi G).\label{t3.3}
\end{eqnarray}
The main objective of this paper is to demonstrate that the cosmological model described above can be seen to be quantum theory-friendly. That is to say it can be physically and numerically expressed so that it is not in conflict with quantum theory. The spade work for this has been done in the application papers (\cite{58:gil}, \cite{60:gil}) in which it was shown that firstly the famous {\it cosmological constant problem\/} does not arise in this model and secondly the equally famous {\it cosmological coincidence problem\/} can be removed from the structure of this model, if care is taken in the use of astronomical measurements. The form of the well known cosmological coincidence problem that occurs in this model takes what I call a {\it critical\/} form because it involves the integer $2$ in the result $t_0=2t_c$, where it was thought that $t_0$ should correspond to {\it time now\/} and $t_c$ is a definitely fixed time when the universe's radial acceleration is zero. However, as time now for an observation depends when the observation is made and is in that sense variable, it is difficult to see how $t_0=2t_c$ can be a universal result. Thus it is more rational to define $t_0$ as the definite value of epoch time when the ratio of the quantity of {\it dark energy mass\/} certainly within the universe to the quantity of conserved mass which is for all time within the universe can be thought to have the value $3= \Omega_\Lambda /\Omega_{M_U} $ as identified in terms of the $\Omega$s by astronomical observations at time now which has also commonly been called $t_0$. These astronomical measurements are displayed next.
The accelerating universe astronomical observational workers \cite{01:kmo} give measured values of the three $\Omega s$, and $w_\Lambda$ to be
\begin{eqnarray}
\Omega_{M,0} &=&8\pi G\rho_0/(3 H_0^2)=0.25^{+0.07}_{-0.06}\label{t5}\\
\Omega_{\Lambda,0} &=& \Lambda c^2/(3 H_0^2)=0.75^{+0.06}_{-0.07}\label{t6}\\
\Omega_{k,0} &=& -kc^2/(r_0^2 H_0^2) =0,\ \Rightarrow k= 0\label{t7}\\
\omega_\Lambda &=& P_\Lambda/(c^2 \rho _\Lambda) = -1\pm\approx 0.3.\label{t8}
\end{eqnarray}

I abandon the use of $t_0$ to represent time now or the rather vague time when the measurements were made and represent time now by the symbol $t^\dagger$ which still remains vague but for the purpose of theoretical discussion can taken to be the time of the present moment. The time $t_0$ will be used to represent the much less vague time when the universe passes through the centre value of the measurement range. I intend to re-express the second equalities above but before making that step it necessary to give a more detailed account of what the measurements above mean in their initial form in relation to the form I shall replace them by. This question of meanings and relations follows in a subsection.
\subsection{Cosmological Epoch and Terrestrial Time}  
For clarity I now rewrite the first two equations, (\ref{t5}) and (\ref{t6}) in terms of the time,  $t_0$,
\begin{eqnarray}
\Omega_M (t_0) &=&8\pi G\rho (t_0)/(3 H^2(t_0))=0.25^{+0.07}_{-0.06}\label{t5.1}\\
\Omega_\Lambda (t_0) &=& \Lambda c^2/(3 H^2(t_0))=0.75^{+0.06}_{-0.07},\label{t6.1}
\end{eqnarray}
where the Hubble function given at (\ref{t1.1}) is used. The first equalities in these equations define definite $\Omega (t_0)$ functions of time, whereas the second two equalities say that known functions of $t_0$ lie within definite numerical ranges. 
These second equalities in (\ref{t5.1}) and (\ref{t6.1}) can be usefully rewritten as follows
\begin{eqnarray}
\frac{3\times 0.19}{8\pi G}<\frac{\rho(t_0)}{H^2(t_0)}<\frac{3\times 0.32}{8\pi G}\label{t5.11}\\
\frac{3\times 0.68}{\Lambda c^2}<\frac{1}{H^2(t_0)}<\frac{3\times 0.81}{\Lambda c^2}.\label{t6.11}
\end{eqnarray}
Inverting these equations, we have
\begin{eqnarray}
\frac{8\pi G}{3\times 0.19} >\frac{H^2(t_0)}{\rho(t_0)} >\frac{8\pi G}{3\times 0.32} \label{t5.12}\\
\frac{\Lambda c^2}{3\times 0.68}>H^2(t_0)>\frac{\Lambda c^2}{3\times 0.81}\label{t6.12}
\end{eqnarray}
 and using, (\ref{t2.2}) and (\ref{t3.3}), these equations can be converted to
\begin{eqnarray}
\frac{1}{ 0.19} >\cosh^2 (3c t_0/(2R_\Lambda ) ) >\frac{1}{0.32} \label{t5.13}\\
\frac{1}{ 0.68}>\coth^2 (3c t_0/(2R_\Lambda ))>\frac{ 1}{0.81}.\label{t6.13}
\end{eqnarray}
It follows that these two equations are saying the same thing because
\begin{eqnarray}
\coth^2(3c t_0/2R_\Lambda) =\frac{\cosh^2(3c t_0/(2R_\Lambda))}{\cosh^2(3c t_0/(2R_\Lambda ))-1} \label{t5.14}
.\label{t6.14}
\end{eqnarray}
According to the measurement (\ref{t6}) information the universe will pass through the centre of the $\Omega _\Lambda$ values at some time, $t_0$, say, given by
\begin{eqnarray}
3 c t_0/(2 R_\Lambda) &=& \coth^{-1} ((1/0.75)^{1/2}) \label{t7.1}\\
&=& \coth^{-1} ( 2/3^{1/2})=\cosh^{-1}(2) \label{t7.2}\\
t_0 &=&(2R_\Lambda /3 c) \cosh^{-1}(2). \label{t7.3}
\end{eqnarray}
From (\ref{t7.3}) it is clear that we cannot find a numerical value for the special time $t_0$ unless we can find a numerical value for $R_\Lambda$ and this is equivalent to knowing the numerical value for $\Lambda =(3/R_\Lambda)^{1/2}$. However, in the very unlikely special case of coincidence, when $t^\dagger = t_0$ we can calculate $R_\Lambda$ because the relation first displayed below implies the second and then the third followed by the  definite special case value for $t_0$ at fourth place. 
\begin{eqnarray}
t^\dagger &=&(2R_\Lambda/(3c))coth^{-1}(R_\Lambda H^\dagger/c) \label{t7.11}\\
&=& t_0 = (2R_\Lambda/(3c))cosh^{-1}(2) \label{t7.21}\\
R_\Lambda &=&  2c/(3^{1/2} H^\dagger)\approx 1.48353 \times 10^{26}. \label{t7.31}\\
t_0 &=& 4.35\times 10^{17}\ s\approx 4.756\times 10^{11}\ yr.\label{t7.32}
\end{eqnarray}
In fact, the value of $t_0$ given at (\ref{t7.32}) is the theoretically given value mentioned earlier for the time when $\Omega_\Lambda /\Omega_{M_U}=3 $, an event that occurs inevitably, a result independent of measurement.
From the third equality above and $t_0\not = t^\dagger$  we get the general result
\begin{eqnarray}
R_\Lambda =3 c t_0 \coth^{-1}(3^{1/2})=(3 c t_0/2)\cosh^{-1} (2).\label{t7.33}
\end{eqnarray}
There is an important lesson from the general result (\ref{t7.33}) which is that if $t_0$ is determined in value then so is $R_\Lambda =(3/\Lambda)^{1/2}$ or $\Lambda$ and visa versa.
The time $ t_{0,min}$ when the time $t_0$ is at the lower measurement value and the time $ t_{0,max}$ when the time $t_0$  is at the higher measurement value are given, using (\ref{t6.13}) and the fact that in general $R_\Lambda$ is to be determined, by
\begin{eqnarray}
t_{0,min}&=& \left(\frac{2R_\Lambda}{3 c}\right)\coth^{-1}((1/0.81)^{\frac{1}{2}})\approx 3.8634\times 10^{17}\ s\label{t7.51}\\
t_{0,max}&=& \left(\frac{2R_\Lambda }{3 c}\right)\coth^{-1}((1/0.68)^{\frac{1}{2}})\approx 4.8568\times 10^{17}\ s\label{t61}\\
t_{0,mean}&=& ( t_{0,min} + t_{0,max})/2 \approx 4.3601\times 10^{17}\ s\label{t9}\\
t_{0,mean} - t_0 &\approx & 10^{15}\ s\approx  3.17\times 10^{7}\ yr\label{t9.1}\\
t_{0,mean}/t_0&\approx & 1.00232.\label{t9.11}
\end{eqnarray}
Thus the length of the time range between which time , $t_0$, viewed as a variable over the measurement range, can be expected to be found is given by
\begin{eqnarray}
t_{0,max}- t_{0,min}&=& 0.9934\times 10^{17}\ s\approx 3.15\times 10^{9}\ yr\label{t7.7l}\\
t_{0,max}/ t_{0,min}&=&4.8568/3.863\approx 1.2572 \label{t7.8}\\
t_{0,max}/ t_0&=&4.8568/4.756\approx 1.02119.\label{t7.9}
\end{eqnarray}
The quantities $ t_{0,min}$ and $ t_{0,max}$ are here taken to be the lower and upper time limits associated with the measurements of the omegas given at equations (\ref{t5.1}) and (\ref{t6.1}). In all the evaluations of times above, $R_\Lambda$ has been given the special case coincidence value. In the non-coincidence general case it would have a value different from this but this value could only be determined by some new or other experimental procedure. The coincidence value has been used just to give some idea of the various bounds of the quantities involved quantities.
From these equations assumed to hold at a conceptual time, $t_0$, when the universe passes through the centre value of the measurement ranges, we get the exact formulae,
\begin{eqnarray}
t_0&=& (2 R_\Lambda /(3 c))\cosh^{-1}(2)\label{tt12}\\
R_\Lambda &=& 3ct_0/(2 \cosh^{-1}(2))\label{tt13}\\
t_c &=& (t_0 \cosh ^{-1}(2))\coth^{-1}(3^{1/2})\label{tt14}\\
t_0/t_c &=& \cosh^{-1}(2)/\coth^{-1}(3^{1/2})=2.\label{tt15}
\end{eqnarray}
Having found $R_\Lambda$ in terms of $t_0$, this value of $R_\Lambda$ can be substituted into the formula for Hubble's constant, (\ref{tt16}),  to find the value of the {\it time now\/}, $t^\dagger$. 
\begin{eqnarray}
H(t^\dagger) &=& (c/R_\Lambda) \coth(3 c t^\dagger /(2 R_\Lambda )) \label{tt16}\\
t^\dagger &=& (2R_\Lambda /(3 c) )\coth^{-1} (R_\Lambda H^\dagger/c)\label{tt17}\\
&=& \left(   \frac{t_0}{\cosh^{-1}(2)}  \right)\left(  \coth^{-1}\left(  \frac{3t_0H^\dagger }{2\cosh^{-1}(2)}\right)\right)\label{tt18}\\
&=& \left(   \frac{2t_c}{\cosh^{-1}(2)}  \right)\left(  \coth^{-1}\left(  \frac{6t_c H^\dagger }{2\cosh^{-1}(2)}\right)\right),\label{tt19}
\end{eqnarray}
where $H^\dagger= H(t^\dagger)$ is the present day measured value of Hubble's constant.
Equations (\ref{tt18}) or (\ref{tt19}) is essentially the solution to the coincidence problem. If we write (\ref{tt19}) in the form
\begin{eqnarray}
t^\dagger /t_c &=& \left(\frac{2}{\cosh^{-1}(2)}  \right)\left(\coth^{-1}\left(\frac{6t_c H^\dagger}
{2\cosh^{-1}(2)}\right)\right)\label{tt20}\\
t^\dagger /t_c &=& 2f(2 t_c),\label{tt21}
\end{eqnarray}
where $f(2t_c)$  gives the deviation of the ratio $t^\dagger /t_0$ from the value unity and removes the degeneracy. Expressed in another way it is the multiplicative function that breaks the coincidence at (\ref{tt15}) and converts the integer $2$ to a much less notable non integral value. However, we can give the formulae (\ref{tt20}) and (\ref{tt21}) together an interpretation in terms of the uncertainties of the measurement process.
This is achieved by defining the measurement {\it deviation\/} function $d_{meas}(t_0)$ as follows,
\begin{eqnarray}
d_{meas}(t_0)&=& t^\dagger /t_0  - f(t_0) \label{tt22}\\
f(t_0) &=& \left(\frac{1}{\cosh^{-1}(2)}  \right)\left(\coth^{-1}\left(  \frac{3t_0 H^\dagger }{2\cosh^{-1}(2)}\right)\right) .\label{tt23}
\end{eqnarray} The function (\ref{tt22}) is a dimensionless measure of how much the central $\Omega$ values from astronomy assumed to have occurred at $t_0$ differ from the time now measurement from the Hubble variable quantity $H(t^\dagger)$ taken at time now, $t^\dagger$. It is sufficient to assume that the event at $t_0$ is still yet to occur, $t_0 > t^\dagger$, then we see that the function $d_{meas}$ passes through zero when the full degeneracy holds at $t_0 = t^\dagger$ and it has a maximum at $t_0\approx 0.643\times 10^{18} s$ when $t^\dagger$ and $t_0 $ assume the approximate maximum deviation, $0.17$.
When $t_0=0.643\times 10^{18}$, $t^\dagger$ can be assumed constant at the coincidence value $4.34467 \times 10^{17}$  so that the maximum deviation times ratio is $t^\dagger /t_0 \approx 0.43467 / 0.643 \approx 0.6757$ or
\begin{eqnarray}
t^\dagger =0.6757t_0.\label{tt24}
\end{eqnarray}
It follows that $t^\dagger$, the time now value, can vary from $t_0$ down to a value of $t^\dagger \approx 0.6757t_0 =1.3514t_c$. Thus the coincidence is decisively removed with $t^\dagger \not = t_0 = 2 t_c$. This calculation is based on the assumption that the true physical value of the Hubble function at time $t^\dagger$, $H(t^\dagger)$, is the measured {\it central\/} value $H^\dagger$ used in the formulae (\ref{tt18}) and (\ref{tt19}). However, $H(t^\dagger)$ could have any value in its measurement range which according to {\it W. Freedman\/}, \cite{61:free}, is 
\begin{eqnarray}
72\pm 8\  Kms^{-1}Mpc^{-1}\approx (2.33\pm 0.25)\times 10^{-18}\ s^{-1}.\label{tt25}
\end{eqnarray}

This implies that the quantity $H^\dagger$ used in the formulae (\ref{tt18}) and (\ref{tt19}) could have values in inverse seconds in the range

 \begin{eqnarray}
H^\dagger _{min} = 2.07\times 10^{-18} < H^\dagger < 2.58 \times 10^{-18} = H^\dagger _{max}.\label{tt26}
\end{eqnarray}
the quantities $H^\dagger _{min}$ and $H^\dagger _{max}$ defined at equation (\ref{tt26}) can then be used to produce two further versions of equations (\ref{tt22}) and (\ref{tt23}) referring here to the limit end points of the Hubble measurement range
\section{Coincidence Deviations Diagram}
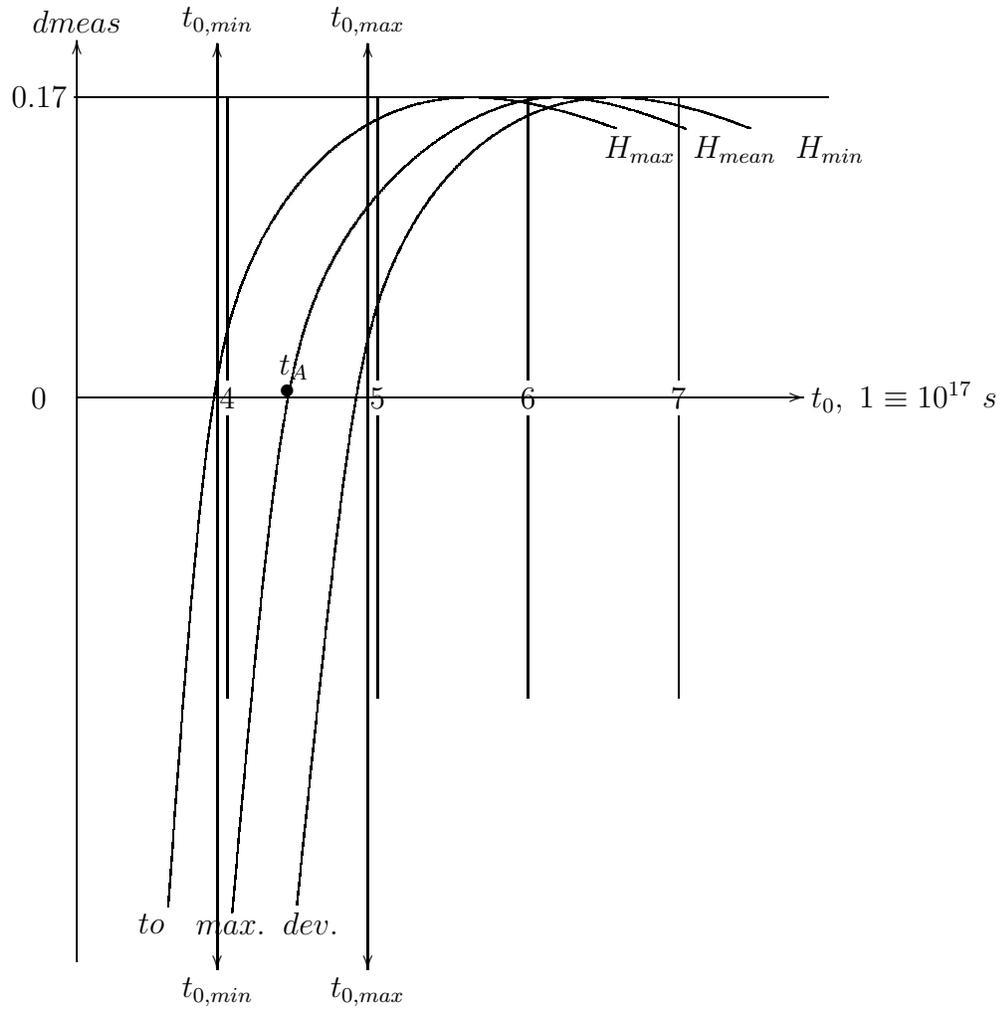
\xy
\POS(50,-90)*+{Max.\ neg.\ deviation\ where\  curve\  H_{min}\rightarrow dev.\  meets\  abscissa \ t_{0,min}}
\POS(50,-95)*+{Max.\ pos.\ deviation\ where\  curve\  H_{max}\rightarrow to\  meets\  abscissa\ t_{0,max}}
\POS(50,-100)*+{t_A\ astronomers'\  consensus\ universe's\ age,\ dmeas\ within\  bullet}
\POS(-5,0)*+{0}
\POS(0,0)*+{}
\ar @{->} (110,0)*+{t_0,\  1\equiv 10^{17}\ s}
\ar @{->} (0,50)*+{dmeas}
\ar @{-} (0,-75)*+{}
\POS(20,0)*+{4}
\ar @{-} (20,40)*+{}
\ar @{-} (20,-40)*+{}
\POS(18.7,0)*+{}
\ar @{->} (18.7,50)*+{t_{0,min}}
\ar @{->} (18.7,-79)*+{t_{0,min}}
\POS(38.7,0)*+{}
\ar @{->} (38.7,50)*+{t_{0,max}}
\ar @{->} (38.7,-79)*+{t_{0,max}}
\POS(40,0)*+{5}
\ar @{-} (40,40)*+{}
\ar @{-} (40,-40)*+{}
\POS(28.9,4)*+{t_A}
\ar @{.} (28.0,1)*+{}
\POS(60,0)*+{6}
\ar @{-} (60,40)*+{}
\ar @{-} (60,-40)*+{}
\POS(28.0,0.9)*+{\bullet}
\POS(80,0)*+{7}
\ar @{-} (80,40)*+{}
\ar @{-} (80,-40)*+{}
\POS(100,40)*+{}
\ar @{-} (-5,40)*+{0.17}
\POS(96,33)*+{\ \ \ \ \ \ H_{min}};\POS(29,-70)*+{\ \ \ dev.}
**\crv{(81,40)&(61,43)&(38.5,20)&(35.2,-10)}
\POS(79,33)*+{H_{max}\ \ \ \ \ \ };\POS(12,-70)*+{to\ \ \ }
**\crv{(62,40)&(42,43)&(19.5,20)&(16.2,-10)}
\POS(87.5,33)*+{H_{mean}};\POS(20.5,-70.5)*+{max.}
**\crv{(71.5,40)&(56.5,43)&(29.5,20)&(25.7,-10)}
\endxy 
\begin{eqnarray}
d_{meas,min}(t_0)&=& t^\dagger_{min} /t_0  - f_{min}(t_0) \label{tt27}\\
f_{min}(t_0) &=& \left(\frac{1}{\cosh^{-1}(2)}  \right)\left(\coth^{-1}\left(  \frac{3t_0 H^\dagger _{min} }{2\cosh^{-1}(2)}\right)\right) \label{tt28}\\
t^\dagger_{min} &=& 4 cosh^{-1}(2)/(3\times3^{1/2} H^\dagger _{min})\label{tt29}\\
d_{meas,max}(t_0)&=& t^\dagger _{max} /t_0  - f_{max}(t_0) \label{tt30}\\
f_{max}(t_0) &=& \left(\frac{1}{\cosh^{-1}(2)}  \right)\left(\coth^{-1}\left(  \frac{3t_0 H^\dagger _{max} }{2\cosh^{-1}(2)}\right)\right) \label{tt31}\\
t^\dagger_{max} &=& 4 \cosh^{-1}(2)/(3\times3^{1/2} H^\dagger _{max})\label{tt32}
\end{eqnarray}
All three of the function $d_{meas,min}(t_0)$, $d_{meas}(t_0)$ and $ d_{meas,max}(t_0)$ have a maximum deviation value a $\approx 0.17$ at a value of $t_0$ appropriate for the function in question. However, the range of the variable $t_0$ in these functions is constrained by the lower and upper limits for the variable $t_0$, $t_{0,min}$ and $t_{0,max}$ given earlier, (\ref{t7.51}) and (\ref{t61}). The maximum value of the deviations with the measurements available lie outside the limits. Thus the actual positive deviations that seem to be possible under these constraints are reduced numerically to parts of the curves within the constraints. Negative deviations occur of about $-0.34$, and $-0.51$ which refer to the case when $t_0$ occurs before $t^\dagger$, a type of situation not discussed in previous work.
The results obtained above can now be used to discuss in section (\ref{sec-tnp}) what I shall call the {\it time-now problem\/}.  
\section{Time-Now Problem}
\setcounter{equation}{0}
\label{sec-tnp}
The time quantities $t_0$, time when the Omegas were measured, and $t^\dagger$, time now, have conventionally been taken to be the same physical quantity with the same numerical value. As I have shown above, this led to the so called {\it cosmological coincidence problem\/}. Both these time quantities have been regarded as representing {\it time now\/} and this in spite of the fact that the two separate measurements involved occurred at distinctly {\it different\/} terestrial times. Consequently, the explanation for the conceptual mistake involved in generating the cosmological coincidence problem is worth further discussion as it throws light on the nature of and reason for the coincidence difficulty that for years has seemed to be so intractable. I suggest the problem arises from the nature of {\it epoch time keeping\/} in cosmology which involves enormously large numerical values such as $10^{10}$ years and also from the relation between {\it epoch\/} time and {\it terrestrial\/} time. It seems that the equality of these two, very distinct time keeping systems, has been taken for granted. We can analyse these issues using the information about the measurements and the times when they occurred using the two quantities $\Omega_\Lambda(t)$ and $H(t)$. In the theoretical cosmology structure that I have been using, both of the quantities are explicit functions of the time variable $t$ and this time variable is the epoch time that occurs in the Friedman theory and derives from Einstein's field equations. However, the astronomers refer to the closely related object pair $\Omega _{\Lambda, 0}$ and $H_0$ as the measured values of these quantities at the time of measurement or {\it roughly\/} speaking at time now. This is indicated by the zero subscript and essentially  means {\it time now\/}. The time concept that is being used by the astronomers in this context is terrestrial time, time measured by earthbound clocks of some sort or other, adjusted to some running numerical value such as Greenwich mean time. Once this competing time forms situation is recognised a whole collection of uncertainties are released into cosmological theory arena. To analyse this {\it conflicting times\/}  situation between cosmological time and terrestrial time it is necessary to bring some precision into the definitions of the quantities under discussion. In particular, the idea of time now that I have previously denoted by $t^\dagger$ has always been a very vague concept. For one thing the term {\it time now\/} implies some variable like character to the value of its symbol in the sense that time now used today is numerically less than time now when used tomorrow. The reason for this vague use in cosmology is obviously due to the fact that on the cosmological scale very little will have appeared to have changed between today and tomorrow or indeed between 10 years ago or 10 years into the future and this is directly a result of the numerically large numbers associated with time passage in cosmology. This vagueness about the time now idea, I think,  has been a major contributor to the time coincidence problem. For this reason I shall now abandon the use of the symbol $t^\dagger$ as representing the vague time now and firmly only use it to represent the time that is to be found from the measured central value, $H^\dagger$, of the Hubble variable from the theoretical form of that parameter, (\ref{t1.1})
\begin{eqnarray}
H^\dagger=H(t^\dagger)& = &  (c/R_\Lambda ) \coth(3ct^\dagger/(2R_\Lambda ))\label{tt.33}\\
t^\dagger & = & (2 R_\Lambda /(3 c)) \coth^{-1} (R_\Lambda H^\dagger /c ).\label{tt34}
\end{eqnarray}
Thus mathematically nothing has changed. However, I have abandoned the terminology {\it time now\/} for the symbol $t^\dagger$ which is now to be firmly associated only with the cosmological time giving the measured value of $H(t)$. I now  rename $t^\dagger$ as the $2001$-Hubble measurement epoch time or briefly 1Hmeastime. This time is clearly now a fixed constant and indeed one of the very large numbers that occurs for cosmological parameters. In parallel with this new definition for $t^\dagger$, I also rename the cosmological time quantity $t_0$ as $2003$-$\Omega_\Lambda$ measurement epoch time or briefly $3\Omega$meastime and which is also one of the very large numbers that occurs for cosmological parameters. The time $t_0$ is the time when the ratio of conserved positive gravitational mass is exactly one third of the negative gravitational mass, or dark energy mass, within the universe's boundary, when $\Omega _M/\Omega _\Lambda =1/3$. The special importance of the time $t_0$ is that it appears that it should be associated with the central measured value of $\Omega _\Lambda$. See equations (\ref{t7.51}) and (\ref{t61}).
\section{Coincidence Free Universe and Lambda}
\setcounter{equation}{0}
\label{sec-cfu}
We have seen that the effect of using the dust universe solution in the definition for the astronomer's $\Omega _\Lambda (t)$ leads to a very effective means of eliminating the cosmological coincidence problem but results in an inability to calculate the value of Einstein's $\Lambda$ because of complications due to $\Lambda$ in Hubble's function $H(t)$, (\ref{tt.33}) and (\ref{tt44}).
\begin{eqnarray}
H^\dagger=H(t^\dagger)& = &  (c/R_\Lambda ) \coth(3ct^\dagger/(2R_\Lambda))\label{tt.331}\\
t^\dagger & = & (2 R_\Lambda /(3 c)) \coth^{-1} (R_\Lambda H^\dagger /c )\label{tt341}\\
& = & (2 (3/\Lambda)^{1/2}/(3 c)) \coth^{-1} ((3/\Lambda)^{1/2}H^\dagger /c )\label{tt342}
\end{eqnarray}
From equation (\ref{tt342}) it is clear that if $t^\dagger$ were known we could calculate $\Lambda$ and vice versa. However, a measured value of $t^\dagger$ is not given by the measurements so far being used in this work so that $\Lambda$ cannot be calculated. The problem of finding the numerical value of $\Lambda$ rests on finding a measured value for $t^\dagger$. Unfortunately, no direct measurement of the value of $t^\dagger$ seems to be possible at this time in astronomy history. The only way out of this dilemma at the moment seems to me to be to accept the consensus result of many astronomy measurements, extrapolations and speculations that the age of the universe, $t_A$, is approximately $13.7 \times 10^{10}\ yrs \approx 4.320432\times10^{17}\  s$. This value can be assigned to $t^\dagger$ as what might be called a working hypothesis, clearly not correct and very approximate but perhaps the best that can be done at this time. In this spirit, I shall take this value for $t^\dagger$ together with the central measured value of $H(t^\dagger)$ in formula (\ref{tt342}) to enable the finding of a {\it best\/} value for $\Lambda$. It was the identification $t^\dagger = t_0$ that lead to the coincidence problem initially and also made possible the calculation of the initially coincidence value for $\Lambda$, denoted now by the subscript $C$ as $\Lambda _C$. This can all be seen from the following displayed equations by taking $t^\dagger = t_0$. The resulting numerical solution for  $\Lambda _C$ is also displayed and compared with {\it removed\/} coincidence numerical value for $\Lambda$ obtained by taking $t^\dagger = t_A$ at equation (\ref{tt347}). 
\begin{eqnarray}
t_0 &=&(2 (3/\Lambda)^{1/2}/3 c) \cosh^{-1}(2) \label{tt.343}\\
&\approx &  1.52\times 10^8\ yrs    \label{tt.3431}\\
t^\dagger & = & (2 (3/\Lambda)^{1/2}/(3 c)) \coth^{-1} ((3/\Lambda)^{1/2}H^\dagger /c ) \label{tt344}\\
t_A &\approx & 4.320432\times10^{17}\  s \label{tt345}\\
\Lambda_C &=& 1.3631\times 10^{-52} \label{tt346}\\
\Lambda\ \ \! &=& 1.3536\times 10^{-52} \label{tt347}\\
t^\dagger /t_0 &=& \coth^{-1} ((3/\Lambda)^{1/2}H^\dagger /c)/\cosh^{-1} (2)\approx 0.99 \label{tt348}\\
t_0 -t^\dagger &=& 0.01 t_0 \approx 1.52 \times 10^6\  yrs\label{tt3481}
\end{eqnarray}
Equation (\ref{tt348}) is the ratio of $t^\dagger = t_A$ or, essentially the time-now value or age of the universe, to the time, $t_0$,  when the amount of negatively gravitating mass to positively gravitating mass is $3/1$ and the ratio, $ t^\dagger /t_0 $, is less than one. This taken with equation (\ref{tt3481}) has the implication that we have about $1.5$ million years  before the $3/1$ stage in the universe's evolution is reached at time $t_0$.
Einstein's cosmological constant plays an essential and fundamental role in the dust universe model. If $\Lambda$ is put to zero within this model, it ceases to exist because most of the physical quantities involved  become zero. In particular, because the vital time arguments such as $3ct/(2 R_\lambda)\rightarrow 0$. Thus $\Lambda$ is the essential and fundamental constant at the basis of the Friedman dust universe. I claim, ending this section, that this cosmological model, using the value for $\Lambda$ at equation (\ref{tt347}), is both free from the {\it cosmological constant problem\/} and the {\it cosmological coincidence problem\/}. Thus the model is of a suitable form to be used to address the problem of how to quantize cosmology. This development follows in Section~\ref{sec-enlw}  
\section{Universe Expansion as  Non-Linear Wave}
\setcounter{equation}{0}
\label{sec-enlw}
It is convenient here to give a {\it very brief\/} reminder of the structure of Schr\"odinger theory in relation to the Friedman equations.
The two Friedman equations from general relativity and the Schr\"odinger equation from quantum theory  have the following three forms,
\begin{eqnarray}
8\pi G \rho r^2/3 & = & {\dot r}^2 +(k - \Lambda r^2/3) c^2\label{tu35}\\
-8 \pi GP r/c^2 & = & 2 \ddot r + {\dot r}^2/r +(k/r -\Lambda r) c^2 \label{tu36}\\
i\hbar \frac {\partial \Psi ({\bf r},t)}{\partial t} &=& -\frac{\hbar ^2}{2m} \nabla^{2} \Psi({
\bf r},t) +V({\bf r})\Psi({\bf r},t) \label{tu37}\\
E_n \Psi_n ({\bf r},t)&=& i\hbar \frac {\partial \Psi_n ({\bf r},t)}{\partial t}\label{tv37}\\ 
\nabla &=& {\bf i} \partial/\partial x +{\bf j} \partial /\partial y +{\bf k} \partial /\partial z \label{tu37.11}\\
\rho _Q({\bf r},t) &=& \Psi ({\bf r},t)\Psi ^* ({\bf r},t) \label{tu37.1}\\
\Psi({\bf r},t) &=& \sum_n \int c_n\Psi_n({\bf r},t).\label{tu37.2}
\end{eqnarray}
At equation (\ref{tu37.1}) the usual definition of the quantum {\it probability\/} density, $\rho _Q({\bf r},t)$, is given as the product of the wave function for the state $\Psi({\bf r},t)$ with its complex conjugate $\Psi^*({\bf r},t)$. At (\ref{tu37.2}), is given the {\it crucially\/} important equation from quantum theory that is called the principle of linear superposition. It is here written in a somewhat symbolic form but in fact means that the solution of the schr\"odinger equation at (\ref{tu37}) can be expressed as a discrete sum or integral or both over constants $c_n$ times the energy eigen-functions, given by equation (\ref{tv37}), of the Schr\"odinger equation. In order to attempt to get at a quantum theory of cosmology or gravity it is reasonable to attempt to convert the two Friedman equations (\ref{tu35}) and (\ref{tu36}) into a form similar to the Schr\"odinger equation (\ref{tu37}). The possibility of such a transformation would seem to depend on what the equations from cosmology and the equation from quantum mechanics have in common. There are two Friedman equations and in fact there are two Schr\"odinger equations because a real part and an imaginary part are added together to make the complex function $ \Psi ({\bf r},t)$ of the space and time variables. Both sets have an important density function associated with them, the mass density $\rho (t)$ in the Friedman set, and the probability density, $\rho _Q({\bf r},t)$,  for the quantum set. Other than these two features they seem to have little in common and the likelihood of finding a conversion from one set to the other seems remote in the extreme. Such a conclusion is reinforced by the recognition of the well known fact that General relativity is a non-linear theory, a non-linearity it transfers to its offspring the Friedman equations, whereas the schr\"odinger equation is linear. However, I shall show in the following pages that this very non-linearity of the Friedman equations can be precisely evaluated and expressed relative to the linearity of the Schr\"odinger equation. This step leads to the formulation of a non-linear wave theory for gravitational waves in contrast with existing gravitational wave theory. The existing theory of gravitational waves usually involves a crude linearisation of general relativity describing waves that have so far not been detected. There is an exception to this by what is called an {\it exact\/} gravitational {\it sandwich\/} wave of rather unconvincing theory that has also not been detected, see Rindler, page $284$, \cite{03:rind}. The waves that are to be described in the following pages have been detected and in fact constitute the expanding or contracting universe structure.
The non-liearity of the Friedman set can be expressed relative to the linearity of the Schr\"odinger set, (\ref{tu37}), symbolically as follows,
\begin{eqnarray}
f(t) = \frac{\sum_j \int n_{f,j}(t)}{\sum_j \int d_{f, j}(t)},\label{tu37.3}
\end{eqnarray}
the space variation of the Schr\"odinger equation does not occur because the cosmology structure does not depend on local space variations at fixed time. The $n$ and $d$ functions refer to the placement of the superposed terms, either numerator or denominator. The function, $f(t)$, just represents what function is being analysed by the non-linear form. I shall call equation (\ref{tu37.3}) the {\it bilinear superposition principle\/} for cosmology. It shows clearly that the non-linearity of cosmology involves the ratio of two linear suppositions of the quantum mechanics type. This principle will be derived in the following, while showing that it allows the contraction and expansion of the universe to be represented as a non-linear standing spherical gravitational wave of time varying radius. The next step in developing this formalism will be to analyse the main mass densities that occur in this model by showing that they can be expressed in the form of the  bilinear superposition principle, (\ref{tu37.3}). 
Two basic  {\it positive everywhere and for all time\/} mass densities have been identified in the development of this cosmological theory. They are the mass density of positive gravitational mass $\rho(t)$ and the mass density of  {\it negative gravitational mass\/} $\rho^\dagger_\Lambda$ which is also everywhere and for all time constant. A third mass density, $\rho_G = \rho (t) -\rho ^\dagger_\Lambda$ is the density of positive gravitational mass relative to negative gravitational mass or alternatively it can be called the gravitational weighted mass density, $\rho _G (t) = (G_+\rho (t) + G_- \rho^\dagger_\Lambda )/G, \ G_-=-G,\  G_+=+G$. This last mass density does become {\it negative\/} for some epoch times. The definition for $\rho (t)$ is,
\begin{eqnarray}
\rho (t) & = & (3/(8\pi G))(c/(R_\Lambda)^2\sinh ^{-2} (3ct/(2 R_\Lambda ))\label{tt.35}\\
& = & A\sinh ^{-2} (3ct/(2 R_\Lambda )) \label{tt36}\\
A & = &(3/(8\pi G))(c/R_\Lambda)^2\label{tt37}\\
R_\Lambda & = & (3/\Lambda)^{1/2},\label{tt37.1}
\end{eqnarray}
where the constant $A$ is introduced as a convenient simplification at formula (\ref{tt36}) and $\Lambda$ is Einstein's cosmological constant. Inspection of the formula for $\rho (t)$ reveals it to be a positive function because the $\sinh$ appears squared. It is also time symmetric, $t\rightarrow -t$, leaves its value unchanged also because of the square. Thus it can be replaced by placing a modulus sign about the time variable as
\begin{eqnarray}
\rho (t)  = A\sinh ^{-2} (3c|t|/(2 R_\Lambda )) ,\label{tt38}
\end{eqnarray}
and this will apply for all time, $-\infty < t < +\infty$. This means that we can use the inverse Fourier transform relation,
\begin{eqnarray}
\exp (-\omega_\Lambda |t|))&=&(2\omega_\Lambda /\pi )\int_0^\infty cos (\omega t)d\omega /(\omega_\Lambda ^2 +\omega ^2),\label{tt39}\\
\exp (-\omega_\Lambda |t|))&=&  (2/\pi )\int_0^\infty cos (\omega_\Lambda st)ds/(1 +s^2),\label{tt40}\\
s&=&\omega /\omega_\Lambda,\label{tt41}
\end{eqnarray}
where $\omega_\Lambda$ is a constant and more conveniently in the second form where $s$ is a none-dimensional dummy to express suitable functions of  decreasing with $|t|\rightarrow \infty$ quantities as functions of integrals over the oscillatory  quantity,  $\cos (\omega t) = \cos (\omega |t|)$. $\rho (t)$ is a suitable function because
\begin{eqnarray}
\rho (t) & = &A \sinh ^{-2} (3ct/(2 R_\Lambda ))\label{tt.42}\\
& = & A((\exp (\omega_\Lambda |t|)-\exp (-\omega_\Lambda|t|))/2)^{-2} \label{tt43}\\
& = & A\left(\frac{  1-\exp (-2\omega_\Lambda|t|) }{ 2\exp (-\omega_\Lambda |t| ) }\right )^{-2} \label{tt44}
\end{eqnarray}
\begin{eqnarray}
& = & A\left(\frac{1-2\exp (-2\omega_\Lambda|t|)+\exp (-4\omega_\Lambda|t|)}{4\exp (-2\omega_\Lambda |t|) }\right)^{-1} \label{tt45}\\
& = & \frac{4A\exp (-2\omega_\Lambda |t|)}{1-2\exp (-2\omega_\Lambda|t|)+\exp (-4\omega_\Lambda|t|)} \label{tt46}\\
\rho^{1/2} (t) & = & \frac{2A^{1/2}\exp (-\omega_\Lambda |t|)}{1-\exp (-2\omega_\Lambda|t|)} \label{tt46.1}\\
A & = &(3/(8\pi G))(c/R_\Lambda)^2\label{tt47}\\
\omega _\Lambda &=& 3c/(2 R_\Lambda)\approx 3.0312 \times 10^{-18}\ cs^{-1}. \label{tt48}
 \end{eqnarray}
We note from equation (\ref{tt40}) that by giving the constant $\omega_\Lambda $ the value zero the general Fourier transform result below follows
\begin{eqnarray}
1&=& (2/\pi )\int_0^\infty ds/(1 +s^2).\label{tt49}
\end{eqnarray}
 Thus all the terms in the fraction at equation (\ref{tt46}) including the zero frequency unit term in the denominator can be expressed as integrals over the dummy variable $s$. Both the numerator and the denominator can be expressed as a superposition of oscillatory cosines or unity, the unit term being included using equation (\ref{tt49}). It should be noted that once the inverse Fourier transform involving the $\cos(n\omega_\Lambda |t|)$ functions are accepted, all such functions can be replaced with $\cos(n\omega_\Lambda t) $ because cosines are even functions anyway. The numerator and denominator of the fraction involved are as follows, 
\begin{eqnarray}
N_\rho (t)&=& 4A\exp (-2\omega_\Lambda |t|)\label{tt50}\\
&=& \frac{8A}{\pi} \int_0^\infty \frac{\cos (2\omega_\Lambda st)}{1 +s^2}ds \label{tt51}\\
n_\rho (t,s)&=&\frac{8A}{\pi} \left(\frac{\cos (2\omega_\Lambda st)}{1 +s^2}\right) \label{tt50.1}\\
N_\rho (t)&=& \int_0^\infty n_\rho (t,s)ds\label{tt50.2}\\ 
D_\rho (t)&=&1-2\exp ( -2\omega_\Lambda|t|)+\exp (-4\omega_\Lambda|t|).\label{tt52}\\
&=&\frac{2}{\pi}\int_0^\infty \frac {1-2 \cos (2\omega_\Lambda st)+  \cos (4\omega_\Lambda st)}{1+s^2}ds\label{tt53}
\end{eqnarray}
\begin{eqnarray}
d_f(t,s) &=&\frac{2}{\pi}\left(\frac {1-2 \cos (2\omega_\Lambda st)+  \cos (4\omega_\Lambda st)}{1+s^2}\right)\label{tt53.1}\\
\rho (t) &=& \frac{N_\rho (t)}{D_\rho (t)}=\frac{ \int_0^\infty n_\rho (t,s)ds }{\int_0^\infty d_\rho (t,s)ds } .\label{tt54}
\end{eqnarray}
Before discussing the significance of formulae (\ref{tt51}), (\ref{tt53}) and (\ref{tt54}), it is useful to consider the integral form for the dark energy density, $ \rho^\dagger _\Lambda = (3/(4\pi G))(c/R_\Lambda)^2$. As this is a constant it can be written in the form
\begin{eqnarray}
\rho^\dagger _\Lambda = \frac{4A}{\pi }\int_0^\infty\frac{ds}{1 + s^2} =2 \rho _\Lambda.\label{tt55}
\end{eqnarray}
This can be used to find oscillatory based form for the gravitationally weighted mass density $\rho_G$,
\begin{eqnarray}
\rho _G &=& \rho (t) -\rho ^\dagger _\Lambda\label{tt56}\\
&=& - 2A\frac{1-4\exp( -2\omega_\Lambda|t|) +\exp (-4\omega_\Lambda|t|)}{ (1-\exp (-2\omega_\Lambda|t|))^{2}}.\label{tt57}
\end{eqnarray}
Thus $\rho _G (t)$ can be expressed in terms of the ratio of numerator and denominator given by
\begin{eqnarray}
N_G(t)&=& - 2A(1-4\exp( -2\omega_\Lambda|t|) +\exp (-4\omega_\Lambda|t|))\label{tt58}\\
&=& - \frac{4A}{\pi} \int_0^\infty \frac {1-4 \cos (2\omega_\Lambda st)+  \cos (4\omega_\Lambda st)}{1+s^2}ds \label{tt59}\\
n_G (t,s)&=& - \frac{4A}{\pi}\left(\frac {1-4 \cos (2\omega_\Lambda st)+  \cos (4\omega_\Lambda st)}{1+s^2}\right) \label{tt59.1}\\
N_G (t)&=& \int_0^\infty n_G (t,s)ds\label{tt59.2}\\
D_G (t)&=&(1-\exp (-2\omega_\Lambda|t|))^{2}\label{tt60}\\
&=& \frac{2}{\pi} \int_0^\infty \frac {1-2 \cos (2\omega_\Lambda st)+  \cos (4\omega_\Lambda st)}{1+s^2}ds\label{tt61}
\end{eqnarray}
\begin{eqnarray}
d_G (t,s)&=& \frac{2}{\pi}\left(\frac {1-2 \cos (2\omega_\Lambda st)+  \cos (4\omega_\Lambda st)}{1+s^2}\right)\label{tt61.1}\\
\rho _G (t) &=&\frac{N_G (t)}{D_{G,f} (t)}=\frac{\int_0^\infty n_G (t,s)ds}{\int_0^\infty d_G (t,s) ds} .\label{tt62.2}
\end{eqnarray}
The dark mass dark energy time relational process ratio, $r_{\Lambda,DM}(t)$, can be expanded in oscillatory integrals as
\begin{eqnarray}
r_{\Lambda,DM} (t) &=&\rho^\dagger_\Lambda /\rho (t) = 2 \sinh^{2}(3 c t/(2R_\Lambda)) \label{tt63}\\
&=& 2A/\left( \frac{N_\rho (t)}{D_\rho (t)}\right) =2A\frac{D_\rho (t)}{N_\rho (t)} =2A\frac{\int_0^\infty d_\rho (t,s)ds}{ \int_0^\infty n_\rho (t,s)ds}\label{tt64}\\
&=&\!\frac{\int_0^\infty (1-2 \cos (2\omega_\Lambda st)+  \cos (4\omega_\Lambda st))(1+s^2)^{-1}ds}{2\int_0^\infty \cos (2\omega_\Lambda st)(1 +s^2)^{-1}ds}\label{tt65}
\end{eqnarray} Hubble's function squared is a suitable function for expression as bilinear superposition
\begin{eqnarray}
H^{2}(t) &=& \left(\frac{c}{R_\Lambda}\right)^2 \coth^2 (\omega _\Lambda t) = \left( \left(\frac{c}{R_\Lambda}\right) \frac{ 1+\exp (-2\omega _\Lambda t)}{ 1-\exp (-2\omega _\Lambda t)}   \right )^{2} \label{tt66}\\
N_H(t)&=& \left(\frac{c}{R_\Lambda}\right)^{2} (1+2\exp( -2\omega_\Lambda|t|) +\exp (-4\omega_\Lambda|t|))\label{tt69}
\end{eqnarray}
\begin{eqnarray}
&=& \frac{2}{\pi}\left(\frac{c}{R_\Lambda}\right)^{2}\!  \int_0^\infty \frac {1+2 \cos (2\omega_\Lambda st)+  \cos (4\omega_\Lambda st)}{1+s^2}ds \label{tt70}\\
n_H(t,s)&=& \frac{2}{\pi}\left(\frac{c}{R_\Lambda}\right)^{2} \left(\frac {1+2 \cos (2\omega_\Lambda st)+  \cos (4\omega_\Lambda st)}{1+s^2}\right) \label{tt70.1}
\end{eqnarray}
\begin{eqnarray}
N_H(t)&=& \int_0^\infty n_H(t,s)ds\label{tt70.2}\\
D_H(t)&=&(1-\exp (-2\omega_\Lambda|t|))^{2}\label{tt70.3}\\
&=& \frac{2}{\pi} \int_0^\infty \frac {1-2 \cos (2\omega_\Lambda st)+  \cos (4\omega_\Lambda st)}{1+s^2}ds\label{tt70.4}\\
d_H(t,s)&=& \frac{2}{\pi}\left(\frac {1-2 \cos (2\omega_\Lambda st)+  \cos (4\omega_\Lambda st)}{1+s^2}\right)\label{tt70.5}\\
H^{2} (t) &=&\frac{N_H (t)}{D_H (t)}=\frac{\int_0^\infty n_H(t,s)ds}{\int_0^\infty d_H(t,s) ds} .\label{tt70.6}
\end{eqnarray}
\section{Cosmological Eigen-Functions}
\setcounter{equation}{0}
\label{sec-cef}
Let us consider the bilinear form for $\rho^{1/2} (t)$, the simplest of the bilinear forms,
\begin{eqnarray}
\rho^{1/2} (t) & = & \frac{2A^{1/2}\exp (-\omega_\Lambda |t|)}{1-\exp (-2\omega_\Lambda|t|)} \label{tt70.7}\\
&=& \frac {(4A^{1/2}/\pi)   \int_0^\infty \cos ( \omega_\Lambda st)ds/(1 +s^2)}{(2/\pi) \int_0^\infty (1-\cos (2\omega_\Lambda st)) ds/(1 +s^2) }.\label{tt70.8}
\end{eqnarray}
Because $\cos (\theta) = (\exp (i\theta)+ \exp (-i\theta))/2 $, all the $\cos (\theta)$s in the above formula can be replaced with exponential forms with the result 
\begin{eqnarray}
\rho^{1/2} (t) = 2A^{1/2}\frac{\int_0^\infty (\exp (i\omega_\Lambda st)+ \exp (-i\omega_\Lambda st)) ds/(2 + 2s^2)}{\int_0^\infty (1- (\exp (i2\omega_\Lambda st)- \exp (-i2\omega_\Lambda st))/2) ds/(1 +s^2) }.\label{tt70.9}
\end{eqnarray}
In the following three equations, I introduce two bilinear representations, $\Psi_{nl,\rho} (t)$ and $\Psi^*_{nl,\rho} (t) $ for the total state. They are not independent because one is the complex conjugate of the other and their normal {\it linear superposition\/} represents $\rho^{1/2} (t)$. Their use will be explained later.  
\begin{eqnarray}
\Psi_{nl,\rho ,+} (t) &=& \frac{2A^{1/2}\int_0^\infty \exp (-i\omega_\Lambda st) ds/(2 +2s^2)}{\int_0^\infty (1- (\exp (i2\omega_\Lambda st)- \exp (-i2\omega_\Lambda st))/2) ds/(1 +s^2) }.\label{tt70.91}\\
\Psi^*_{nl,\rho ,-} (t) &=& \frac{2A^{1/2}\int_0^\infty \exp (i\omega_\Lambda st) ds/(2 +2s^2)}{\int_0^\infty (1- (\exp (i2\omega_\Lambda st)- \exp (-i2\omega_\Lambda st))/2) ds/(1 +s^2) }.\label{tt70.92}\\
\rho^{1/2} (t) &=& \Psi_{nl,\rho ,+}(t) +\Psi^*_{nl,\rho,- }(t)= \Psi_{nl,\rho } (t) \label{tt70.93}\\
\rho (t)&=& \Psi_{nl,\rho }(t) \Psi^*_{nl,\rho }(t)  .\label{tt70.94}
\end{eqnarray}
Thus the superposition of $\cos$ forms in these ratios {\it generally\/} can be replaced with the superposition of $\exp$ forms provided a caveat is added to the effect that an appearance of an exponential must always be accompanied with the appearance of its complex conjugate with the same coefficient. With this caveat we can take the exponential forms as representing the fundamental eigen-states of this system, the fundamental ones being taken as the continuous infinite set,
\begin{eqnarray}
\Psi (s,t) = \exp (-i\omega_\Lambda st), \ 0<s<\infty .\label{tt71}
\end{eqnarray}
Thus if these oscillations measured by the angular frequency, $\omega_\Lambda s$, are interpreted as of quantum origin, we can define the cosmological associate energies as
\begin{eqnarray}
E(s) = \hbar \omega_\Lambda s\label{tt72}
\end{eqnarray}
and equation (\ref{tt71}) can be expressed as
\begin{eqnarray}
\Psi (s,t) = \exp (-iE(s)t/\hbar), \ 0<s<\infty \label{tt73}
\end{eqnarray}
with the consequence we get the eignen-value equation equivalent to (\ref{tv37}) repeated below at (\ref{tt75}) for comparison
\begin{eqnarray}
i\hbar \frac {\partial \Psi (s,t)}{\partial t} &=& E(s) \Psi (s,t) \label{tt74}\\
i\hbar \frac {\partial \Psi_n ({\bf r},t)}{\partial t}&=& E_n \Psi_n ({\bf r},t) .\label{tt75} 
\end{eqnarray}
The local variable position vector, ${\bf r}$, does not occur in the cosmology version, (\ref{tt74}), of this equation because all the states involved are uniformly spatially constant and only vary with time. There is also no external potential so that from the cosmology point of view the full Schr\"odinger equation, (\ref{tu37}), {\it essentially\/} reduces to just the energy eigen-value version, equation (\ref{tt74}). The continuous variable $s$ replaces the apparently discrete subscript $n$ but this parameter could under some circumstances also be continuous. The possible range of the dimensionless variable $s$ given at equation (\ref{tt73}) implies that the range of angular frequencies present is given by
\begin{eqnarray}
0< \omega_\Lambda s<\infty .\label{tt76}   
\end{eqnarray}
Inspection of the formulae for the various superposed quantities reveals that the second and fourth harmonics of all the range for, $n\omega_\Lambda s$, $n=2,4$,  frequencies are also simultaneously present so that the range (\ref{tt73}) or (\ref{tt76}) is the complete set as it also includes all the harmonics $n\omega_\Lambda s$ and also includes the dark energy contribution, $n=0$. It is now possible to express $\rho ^{1/2}(t)$ in terms of the bi-linear superposed eigen-values as
\begin{eqnarray}
\rho^{1/2} (t) = \frac{2A^{1/2}\int_0^\infty ( \Psi (-s,t)+ \Psi (s,t)) ds/(1 +s^2)}{\int_0^\infty (1- \Psi (-2s,t)- \Psi (2s,t)) ds/(1 +s^2) }.\label{tt77}
\end{eqnarray}
All the other cosmological functions mentioned above can similarly be expressed in terms of the eigen-value set (\ref{tt76}) and thus it is now possible to express all the essential cosmological structure in terms  of $\rho^{1/2} (t)$ and simultaneously in terms of the $\Psi (n s,t)$ by the following connections
\begin{eqnarray}
\rho (t)&=&(\rho^{1/2}(t))^{2} \label{tt78}\\
\rho^\dagger _\Lambda &=& 2 A \label{tt79}\\
\rho_G (t)&=&\rho (t) -\rho^\dagger _\Lambda \label{tt80}\\
H^{2} (t) &=& (c/R_\Lambda )^{2} ( (\rho (t)/A) +1).\label{tt81}
\end{eqnarray}
The wave motion followed by the dark mass dark energy time relation process seems to me to be particularly basic as it can also be identified, apart from a constant multiplier, $2A/M_U$, as a spherical volume standing wave with front, the boundary of the expanding or contracting universe. It is repeated here for convenience, 
\begin{eqnarray}
r_{\Lambda,DM} (t) &=&\rho^\dagger_\Lambda /\rho (t)= (2A/M_u)V_U (t)\label{tt82}\\
&=&\!\frac{\int_0^\infty (1-2 \cos (2\omega_\Lambda st)+  \cos (4\omega_\Lambda st))(1+s^2)^{-1}ds}{2\int_0^\infty \cos (2\omega_\Lambda st)(1 +s^2)^{-1}ds}\label{tt83}
\end{eqnarray}
$V_U (t)$ being the volume of the universe at time $t$. This same type of gravitational wave motion also applies to the evolution of smaller sub-volumes of mass within the universe with a time origin different from the universe's singular epoch time, $t=0$. \cite{62:gil}.   
This cosmology theory is thus directly derivable from the Schr\"odinger like eigen-equation (\ref{tt74}), if the linear superposition principle of quantum theory is replaced with the {\it bilinear superposition principle.\/} The function $r_{\Lambda,DM} (t)$ is the ratio of the amount of {\it negatively gravitating positive mass within\/} the universe to the amount of {\it positively gravitating positive mass within\/} the universe as a function of epoch time, t. At equation (\ref{tt82}), we see that this ratio is proportional to the volume of the universe and as a function of time it has the same shape as does the volume of the universe as a function of time. This shape is near infinite radius at $t \approx -\infty$ to zero radius at the singularity at $t=0$ on to near infinite radius at $t \approx +\infty$. Thus the wave motion that correspond to this time shape can be thought of a converging to zero spherical wave front at negative times to a diverging from zero spherical wave front at positive times. On the other hand, the reciprocal of this ratio, the ratio of positive gravitating mass within the universe to the negative gravitating  mass within the universe can be seen to be a spherical standing wave with centre fixed at $r=0$ with time varying radius, infinite at $t=0$ and zero at $t=\pm \infty$. Possibly the best and most instructive image for the wave motion comes from studying the wave form of,
\begin{eqnarray}
\rho_G (t)/\rho^\dagger _\Lambda = \rho (t)/\rho^\dagger _\Lambda -1, \label{tt83.1}
\end{eqnarray}
from which it can be inferred that the motion is {\it not\/} waves of density, but rather waves of the gravity  associated with the whole body of the universe. The nodes for this motion occur at $t=\pm t_c$ with the value zero, when the acceleration changes sign through zero, and the antinodes occur at $t=0$ and $t=\pm\infty$ with the values infinity and $-1$ respectively. The waves appear to be waves in the gravitationally polarised {\it aether\/} (\cite{63:gil}), formed from the positively gravitating universe's mass, $M_U$, and the negatively gravitating mass of Einstein's dark energy.
However, whatever visualisation is chosen, it remains a {\it non-linear wave process\/} formed by the bilinear superposition of eigen-solutions of the Schr\"odinger like equation (\ref{tt74}). Again it should be noted that Einstein's cosmological constant is central to the representation of gravitational waves used in this theory because all the frequencies for these waves depend essentially on it. From equation (\ref{tt76}) it follows that this special frequency range would not exist if $\omega_\Lambda = (3\Lambda)^{1/2}c/2\rightarrow 0$, as it would if $\Lambda \rightarrow 0$.
Returning to the question of finding the full Schr\"odinger equation representation for the Friedman dust universe, an objective that can be achieved simply by operating on the non-linear cosmology state function ,$\Psi_{nl,\rho } (t) $ (\ref{tt70.91}), with the quantum energy operator $ \hbar\partial/\partial t$ without its usual imaginary unit, $i$. The result is
\begin{eqnarray}
\hbar\partial \Psi_{nl,\rho} (t)  /\partial t &=& (V_C (t) ) \Psi_{nl,\rho } (t)
\label{tt83.2}\\
V_C (t)&=& -(3\hbar/2)H (t).\label{tt83.3}
\end{eqnarray}
Equation (\ref{tt83.2}) is, not surprisingly, a non-linear Schr\"odinger  equation with a time dependent {\it feedback\/} potential function given by (\ref{tt83.3}) and  with no dependence on a local vector position, ${\bf r}$, features also not surprising. It is essentially the {\it quantum\/} description of the influence of dark energy on the conserved mass density $\rho (t)$ for this model. The missing $i$ in the quantum energy operator in equation (\ref{tt83.2}) can be restored by placing an $i$ in the numerator of the $(3 \hbar /2)$ factor of $V_C (t)$. There is no reason in principle why a feed back potential should not be imaginary in a quantum system involving complex wave functions. In fact, the appearance of the $i$ in the potential function is characteristic of standing wave type solutions.
\section{The Probability Density Issue}
\setcounter{equation}{0}
\label{sec-pdi}
In order to bring the cosmological structure into line with the Schr\"odinger quantum structure it is necessary to decide how probabilistic concepts are to be introduced or found in relation to the cosmology eigen-states and their bilinear superposition. We can get a clue to how this might be done from the quantum equation for probability density, $\rho_Q ({\bf r},t)$, (\ref{tu37.1}) which is the Hermitean scalar product of the two amplitudes or wave functions representing this state. In the cosmology context we have seen that composite states formed from bilinear superpositions of eigen-states are needed to play the full state representation role. I have defined such a state representation for the non-linear gravitationally positive mass density, $\rho (t)$ at references (\ref{tt70.91}) to (\ref{tt70.94}), now repeated at (\ref{tt83}) and followed by a repeat of the quantum probability density definition at (\ref{tt85}).
\begin{eqnarray}
\rho (t)&=& \Psi_{nl,\rho }(t) \Psi^*_{nl,\rho }(t)= M_U/V_U (t) \label{tt84}\\
\rho _Q({\bf r},t) &=& \Psi ({\bf r},t)\Psi ^* ({\bf r},t) \label{tt85}\\
\rho _C (t) &=& \rho (t)/M_U = 1/V_U (t) .\label{tt86}
\end{eqnarray}
The function $\rho _C (t)$, the cosmological mass density for positive gravitational mass divided by the mass of the universe, $M_U$, is an obvious contender for representing  cosmological probability density. However it is constant over all positions within the spherical volume of the universe unlike the the quantum probability density, $\rho _Q({\bf r},t) $, that is generally variable over the region under consideration, such variability described by the position vector ${\bf r}$. The quantum probability density can answer the following type of question. Given a spherical region, $S$, in which a particle is certain to be found, what is the probability for finding this particle in some, $s$, sub-region at a fixed time? The answer can be found by integrating the density ${\bf r}$ over the region $s$. In the cosmological, situation with $\rho _C (t)$ not dependent on position the question and answer is somewhat trivial and the answer would be the value of the volume of $s$ divided by volume of $S$, correct but not very interesting and clearly this is a consequence of the cosmology states being spatially uniform in most present theory, a restriction that may well be removed in the future. This type of probability question refers to a fixed time density. However, if we ask the following different question there is a more interesting answer. Given a time variable volume, $V(t)$, in which a particle is certain to be found at any time what is the probability of finding this particle in a fixed valued volume, $v$, at some time $t$. The answer to this question is, $v/V(t)$ and clearly changes with the changing volume, $V(t)$. This seem to me to be  much more interesting and it fits the cosmological structure of this model when $V(t)$ is taken to be $V_U(t)$.  This type of probability question refers to a changing time density. I shall assume, following the discussion above, that the density, $\rho _C (t) $, defined at equation (\ref{tt84}) and (\ref{tt86}) can represent a probability density of the changing time type. Thus these equations brings the cosmology structure yet closer to the quantum structure at equation (\ref{tt85}).
\section{Conclusions}
The first four sections of this paper are devoted to an expanded discussion of the cosmological coincidence problem. Here the implications of the astronomical measurements of the $\Omega$s and Hubble's constant are examined in more detail than in the previous paper where a solution to this problem was found. It is shown that given the present concensus of the age of the universe it is possible to derive a definite value for Einstein's cosmological constant and at the same time resolve the coincidence problem thus removing a major impediment to the quantization of cosmology. The last sections of this paper are devoted to expressing the contraction and expansion of the universe in terms of gravitational waves. This is achieved by finding the equivalent for cosmology of the quantum principle of linear superposition of states. The principle for cosmology reflects exactly the sense in which cosmology is non-linear in relation to the equivalent linear principle for quantum mechanics and is called {\it bilinear superposition\/}. The bilinear superposition is then used to express the contraction or expansion of the universe as spherical standing wave motion of varying radius. The principle is also used to describe the time dependent relation between dark energy and dark mass as a local radius varying standing wave. It is suggested that this new non-linear gravitational wave motion has been detected in the form of the expanding universe and is thus a reality unlike the usually theorised but not detected linearised wave motion of general relativity.
\newpage
\section{Appendix 4 Abstract}
\centerline{\Large {\bf Solutions of a Cosmological Schr\"odinger}}
\centerline{\Large {\bf Equation for Exact Gravitational Waves}}
\centerline{\Large {\bf based on a Friedman Dust Universe}}
\centerline{\Large {\bf with Einstein's Lambda}}
\vskip0.5cm
Earlier, it was shown that this cosmological model, originally described by the Friedman equations, can be expressed as a solution to a non-linear Schr\"odinger equation. In this appendix, a large collection of solutions to this Schr\"odinger equation are found and discussed in the context of relaxing the uniform mass density condition usually employed in cosmology theory. The surprising result is obtained that this non-linear equation can have its many solutions {\it linearly superposed\/} to obtain solution of the cosmology theory problem of great generality and applicability. 

In the paper $D$, \cite{58:gil}, it was shown that the quantum vacuum polarisation idea can be seen to play a central role in the Friedman dust universe model introduced by the author. In the paper, \cite{64:gil}, it was shown that the Friedman equation structure can be converted into a {\it non-linear\/} Schr\"odinger equation structure. Here, this aspect is further developed by supplementing the solutions to this time only equation with a  dependence on a three dimensional  space position vector, {\bf r}, so that the  equation remains consistent with its cosmological origin. This step then enables finding cosmological models that are not restricted to having a mass density that is certainly time dependent but otherwise remains constant over all {\it three dimensional\/} position space at every definite time.
It is convenient here to repeat a {\it very brief\/} reminder of the structure of Schr\"odinger theory in relation to the Friedman equations.
The two Friedman equations from general relativity and the Schr\"odinger equation from quantum theory  have the following three forms,
\begin{eqnarray}
8\pi G \rho r^2/3 & = & {\dot r}^2 +(k - \Lambda r^2/3) c^2\label{u0}\\
-8 \pi GP r/c^2 & = & 2 \ddot r + {\dot r}^2/r +(k/r -\Lambda r) c^2 \label{u1}\\
i\hbar \frac {\partial \Psi ({\bf r},t)}{\partial t} &=& -\frac{\hbar ^2}{2m} \nabla^{2} \Psi({
\bf r},t) +V({\bf r})\Psi({\bf r},t) \label{u2}\\
E_n \Psi_n ({\bf r},t)&=& i\hbar \frac {\partial \Psi_n ({\bf r},t)}{\partial t}\label{u3}\\ 
\nabla &=& {\bf i} \partial/\partial x +{\bf j} \partial /\partial y +{\bf k} \partial /\partial z \label{u4}\\
\rho _Q({\bf r},t) &=& \Psi ({\bf r},t)\Psi ^* ({\bf r},t) \label{u5}\\
\Psi({\bf r},t) &=& \sum_n \int c_n\Psi_n({\bf r},t).\label{u6}
\end{eqnarray}
The non-linear Schr\"odinger equation that was obtained in reference \cite{64:gil} has the  form
\begin{eqnarray}
i\hbar\partial \Psi_{nl,\rho} (t)  /\partial t &=& (V_C (t) ) \Psi_{nl,\rho }(t)
\label{u7}\\
V_C (t)&=& -(3i\hbar/2)H (t)\label{u8}
\end{eqnarray}
and can be compared with the general linear Schr\"odinger equation at (\ref{u2}). The non-linearity of the cosmological version is indicated by the feedback potential $V_C (t)$, (\ref{u8}) replacing the external potential at (\ref{u2}). The state vector $\Psi_{nl,\rho} (t)$ in the cosmology version initially has no dependence on local position denoted by the three vector, {\bf r}, as in the quantum version, (\ref{u2}). This deficiency will be rectified in the following section.
\section{Position Variable Cosmology Schr\"odinger equation}
\setcounter{equation}{0}
\label{sec-svcs}
Before starting this section, it is necessary to make some remarks about the dimensionality of the usual physical position coordinate vector, {\bf r} =x{\bf i}+ y{\bf j}+ z{\bf k}. This is often taken to have the dimension, $m$, physical length. The relativistic metric used in this theory is of the form
\begin{eqnarray}
 ds^2 = c^2 dt^2 - r^2 (t)(d\grave x^2 + d\grave x^2 + d\grave x^2).\label{u0.1}            
\end{eqnarray}
In this work up to date, I have taken the scale factor $r(t)$ to represent the  {\it physical\/} radius of the universe at epoch time $t$
so that it has the dimension $m$, physical length. If as usual, $c$ has the physical dimensions $ms^{-1}$ and $t$ has the physical dimension,  $s$, then ds in the metric will have the dimension $m$ and  so the vector,  $\grave{\bf r} =\grave x{\bf i}+ \grave y{\bf j}+ \grave z{\bf k}$, will be dimensionless and this is indicated by the above grave accent.
The theory I am working with here is non-linear and attempting to use dimensioned position coordinates can lead to dimensionality chaos. Thus from now on, I shall usually work with the dimensionless position coordinates and use the grave sign to indicate this. Consistent with this policy it is useful it define the dimensionless quantities using the fundamental length $R_\Lambda$ as follows and starting with a dimensionless radius for the universe, $\grave r(t)$,  
\begin{eqnarray}
\grave r(t) &=& r(t)/R_\Lambda\label{u0.2}\\
\grave x &=& x/R_\Lambda\label{u0.3}\\
\grave y &=& y/R_\Lambda\label{u0.4}\\
\grave z &=& z/R_\Lambda.\label{u0.5}                                                
\end{eqnarray}
I shall also use the grave accent to indicate that {\it a function\/} is dimensionless as with $\grave f(r)$.
My strategy in the following work is firstly, to introduce space dependence, $ \grave{\bf r}$, into the cosmological Schr\"odinger equation (\ref{u7}) and then , secondly to show that the introduction of an $\grave{\bf r}$ dependence  can be made consistent with the original Friedman equations structure without damaging their validity as a rigorous solution to Einstein's field equations. Firstly, I rewrite the purely time dependent equation (\ref{u7}) assuming an extra dependence on $\grave{\bf r}$ in the original state vector $\Psi_{nl,\rho }(t)$, while leaving the feedback term unchanged.
\begin{eqnarray}
i\hbar\partial \Psi_{nl,\rho} (t,\grave{\bf r})  /\partial t &=& (V_C (t) ) \Psi_{nl,\rho }(t, \grave{\bf r})
\label{u9}\\
V_C (t)&=& -(3i\hbar/2)H (t).\label{u10}
\end{eqnarray}
The first question that arises is, {\it can this step be done consistently\/}? The answer to this is in the affirmative as can be shown as follows. Rewrite (\ref{u9}) as equation (\ref{u11}) and followed by the time integration at (\ref{u12}) and then inverting the logarithm  at (\ref{u13})
\begin{eqnarray}
\partial \ln\Psi_{nl,\rho} (t,\grave{\bf r})/\partial t &=& -(3/2)H (t)
\label{u11}\\
\ln (\Psi_{nl,\rho} (t,\grave{\bf r})/ \Psi_{nl,\rho} (t_0,\grave{\bf r}))&=& -(3/2)\int_0 ^t H (t^\prime)dt^\prime
\label{u12}\\
\Psi_{nl,\rho} (t,\grave{\bf r}) &=& \Psi_{nl,\rho} (t_0,\grave{\bf r})\exp \left(-\frac{3}{2}\int_{t_0} ^t H (t^\prime )dt^\prime \right) \label{u13}\\
 \Psi_{nl,\rho} (t_0,\grave{\bf r})&=& \Psi_{nl,\rho}(t_0) \grave f(\grave{\bf r}) \label{u13.3}
\end{eqnarray}
Thus introducing a dimensionless function, $\grave{\bf f}$, with $\grave{\bf r}$ dependence presents no problems. It means just multiplying the original time only dependent wave function, $\Psi_{nl,\rho} (t_0)$, with the purely space dependent function, $\grave f(\grave{\bf r})$. This also partially justifies not including any space variation in the Hubble function, $H (t)$. However, this last point will be fully justified when the affect on the  purely time dependent Friedman equations, (\ref{u0}) and (\ref{u1}), is examined in the next paragraph. I should be remarked that the function $\grave f(\grave{\bf r})$ can be a complex valued function in the context of quantum theory wave function structure. This fact will be seen to be useful as the story unfolds.  

The relation of the cosmological Schr\"odinger equation and the Friedman equations clearly has to be mutual consistency. A threat to this consistency is the obvious difference between the purely time dependent mass density function $\rho (t)$ in the Friedman set and the now proposed space time variability through $\grave{\bf r}$ in the Schr\"odinger equation wave function, (\ref{u13}). In using the original Friedman equations, (\ref{u0}) and (\ref{u1}), it has been common practice to assume that $\rho (t)$ is a purely time dependent mass density chosen as a working  approximation to a correct more general time and space dependant version and so rendering difficult mathematics viable though less physically accurate.  This was my starting position when I wrote the first paper, $A$, in this sequence of papers. However, having found the non-linear Schr\"odinger equation (\ref{u7}) it has become clear that the common practice position with regard to $\rho (t)$ needs some modification. My view now is that $\rho (t)$ is a correct quantity in its own right, giving information about the cosmology structure as a global entity. Its definition is repeated below,
\begin{eqnarray}
\rho (t) &=& M_U/ V_U (t) \label{u14}\\
M_U &=& \rho (t) V_U (t), \label{u15}
\end{eqnarray}
where $M_U$ is the total conserved positively gravitational mass of the universe
and $V_U (t)$ is the volume of the universe at epoch time $t$. If $\rho (t)$, does have a definite meaning in its own right and is not just an approximation to a better space dependent version then it can be retained with its self identity as before. This special significance of $\rho (t)$ is effectively retained by keeping it but multiplied by the space dependant contribution as in (\ref{u13.3}). From the existence of a possible true space and time dependent version from Schr\"odinger theory it can be seen that the definition for the  mass, $M_U$, of the universe that appears in (\ref{u14}) with the space dependent density,  should be
\begin{eqnarray}
M_U (t)& =& {R_\Lambda}^3\int\int\int_{V_U(t)}\rho (t,\grave{\bf r})d\grave xd\grave y d\grave z\label{u16}\\
& =& {R_\Lambda}^3\int\int\int_{V_U(t_0)} \rho (t_0,\grave{\bf r})d\grave xd\grave y d\grave z\label{u16.1}\\
&=& M_U =\ a\  constant\label{u17}\\
\rho (t_0,\grave{\bf r})&=&\Psi_{nl,\rho} (t_0,\grave{\bf r})\Psi^*_{nl,\rho} (t_0,\grave{\bf r})=\Psi_{nl,\rho}(t_0) \Psi^*_{nl,\rho}(t_0)\grave f(\grave{\bf r})f^*(\grave{\bf r}) \label{u17.1}\\
&=&\rho (t_0) \grave f(\grave{\bf r})\grave f^*(\grave {\bf r}), \label{u17.2}
\end{eqnarray}
where $V_U (t)$ is the volume of the universe at time $t$, the time dependent spherical volume over which the integration is taken at time $t$ and equations, (\ref{u16}), (\ref{u16.1}) and  (\ref{u17}), holding because the total mass within the universe is a constant over time. In other words $M_U$ is a time conserved quantity or within the universe's  changing boundary, density movement should satisfy the equation of continuity which in the usual coordinates is
\begin{eqnarray}
\partial \rho (t, {\bf r})/\partial t = -\nabla ({\bf v}(t,{\bf r})\rho (t, {\bf r})).\label{u17.3}
\end{eqnarray}
From equations (\ref{u16.1}) and (\ref{u17.2}), we get
\begin{eqnarray}
M_U(t_0) & =& {R_\Lambda}^3\int\int\int_{V_U (t_0)} \rho (t_0,\grave{\bf r})d\grave xd\grave y d\grave z\label{u17.4}\\
& =& \rho (t_0) {R_\Lambda}^3\int\int\int_{V_U (t_0)} \grave f(\grave{\bf r})\grave f^*(\grave{\bf r})d\grave xd\grave y d\grave z .\label{u17.5}
\end{eqnarray}
Thus once the function, $\grave f(\grave{\bf r})$, is chosen, it appears that we  can find the  constant value of the mass of the universe, $M_U$. However, this appearance is deceptive because there is the complication that to get a constant valued numerical value from this equation we have to have a constant valued volume to integrate over while $V_U(t_0)$ depends on $t_0$ and so is in a sense time variable. 
It is necessary to have a value for $M_U$ so that the value of the dimensioned length multiplier
$b= (R_\lambda/c)^{2/3}(2M_U G)^{1/3}$ in the radius of the universe can be considered known,
\begin{eqnarray}
r(t) &=&  b\sinh^{2/3}(\pm 3ct/(2R_\Lambda))\label{U18}\\
b &=&  (R_\Lambda /c)^{2/3} C^{1/3}\label{U19}\\
R_\Lambda &=& (3/\Lambda )^{1/2}\label{U20}\\ 
C &=& 2M_UG.\label{U21}
\end{eqnarray}
Thus we seem to be left with the only options of finding the value of $M_U$ from experiment or just accept that it is an arbitrary dimensioned constant until some alternative route to finding its value is found.
The numerical value of $M_U$ makes no difference to the theoretical structure of the theory, it only effects the numerical value of  Rindler's constant, $C$, and any quantity in which this constant appears as a numerical multiplier which beside $r(t)$ the velocity of expansion $v(t)$ and the acceleration, $a(t)$, are involved. However, {\it importantly\/} for the non-linear Schr\"odinger equation, $H(t)$, does {\it not\/} involve the value of $M_U$,
\begin{eqnarray}
H(t) &=& \dot r(t)/r(t) = (c/R_\Lambda ) \coth(3ct/(2R_\Lambda)).\label{U20.1}
\end{eqnarray}
Because the integral in (\ref{u17.5}) is over the volume of the universe $V_U(t_0)$ which is given by
\begin{eqnarray}
\frac{M_U}{V_U(t_0)}&=&\left(\frac{3}{8\pi G}\right)\left(\frac{c}{ R_\Lambda }\right)^2\sinh^{-2} \left(\frac{3ct_0}{2R_\Lambda}\right)=\rho (t_0)\nonumber\\
&=&\left(\frac{\rho^\dagger_\Lambda }{2}\right)\sinh^{-2} \left(\frac{3ct_0}{2R_\Lambda}\right)=\rho(t_c)= \frac{M_U}{V_U(t_c)}\label{U20.2}\\
M_U(t_0) & =& \rho (t_0) {R_\Lambda}^3\int\int\int_{V_U (t_0)} \grave f(\grave{\bf r})\grave f^*(\grave{\bf r})d\grave xd\grave y d\grave z 
\label{u20.21}\\
M_U(t_0) & =& M_U(t_0) \frac{{R_\Lambda}^3}{V_U(t_0)}\int\int\int_{V_U (t_0)} \grave f(\grave{\bf r})\grave f^*(\grave{\bf r})d\grave xd\grave y d\grave z 
\label{u20.2}\\
1& =& \frac{{R_\Lambda}^3}{V_U(t_0)}\int\int\int_{V_U (t_0)} \grave f(\grave{\bf r})\grave f^*(\grave{\bf r})d\grave xd\grave y d\grave z 
\label{u20.3}
\end{eqnarray}
the relation, (\ref{u20.3}), gives a normalisation condition over physical space on a probability function density  of space position variability,  $\rho_{space}({\bf r})= \grave f(\grave{\bf r})\grave f^*(\grave{\bf r})$, following by cancellation of the mass of the universe $M_U$ in the previous equation, which apparently holds from some definite time, $t_0$, at least. Thus the function $\rho_{space}({\bf r})$ is just what is needed to describe the probability for finding mass at position ${\bf r}$, in the Schr\"odinger equation cosmology context at time, $t_0$. However, consistency demands that equation (\ref{u20.3}) holds, at least, for some specific time $t_0$. Thus we need to check out that such a time exists.
From equation (\ref{U20.2}), we see much that we knew all along but, usefully, we see the value for the volume of the universe at time $t_c$, the time when deceleration  changes to acceleration,  is the obviously very constant value,
\begin{eqnarray}
V_U(t_c) =\frac{M_U}{\rho^\dagger_\lambda} = \left(\frac {4\pi M_UG} {3}\right)\left(\frac{R_\Lambda}{c} \right)^2\label{u20.4} 
\end{eqnarray}
that we need to evaluate the apparently time dependent multiples integrals such as 
\begin{eqnarray}
1& =& \frac{{R_\Lambda}^3}{V_U(t_c)}\int\int\int_{V_U (t_c)} \grave f(\grave{\bf r})\grave f^*(\grave{\bf r})d\grave xd\grave y d\grave z .\label{u20.5}
\end{eqnarray}
Thus we seem very near a prescription for a usable cosmological Schr\"odinger equation. However, given a space dependent solution like (\ref{u13}) it is likely that the part $-\frac{\hbar ^2}{2m} \nabla^{2} \Psi( {
\grave\bf r},t) $ of the quantum version at (\ref{u2}) would occur and this might render the cosmological quantum version not consistent with cosmology. It is by no means certain that such a complication would necessarily occur and not be handleable but certainly it can be avoided by playing safe and imposing the condition on this term as being zero as follows,
\begin{eqnarray}
\frac{\hbar ^2}{2m} \nabla^{2} \Psi(t,\grave{\bf r})=0.\label{U23}
\end{eqnarray}
This implies that the function $\grave f(\grave{\bf r})$ from equation (\ref{u13.3}) also satifies the Laplace equation,
\begin{eqnarray}
\nabla^{2} \grave f(\grave{\bf r})=0.\label{U2.1}
\end{eqnarray}
The {\it Laplace\/} equation has a very large number of solutions. Thus there are many possible space dependent versions for the wave function, $\Psi(t, \grave{\bf r})$. Furthermore, I shall show that in spite of the cosmological Schr\"odinger being non-linear, the many solutions of the Laplace equation can be {\it linearly\/} superposed to produce yet more solutions. Thus although the condition (\ref{U23}) reduces the number of possibilities that might be considered for the space dependent wave function it leaves us more than enough solutions to think about for a very long time. It does have another advantage that could turn out to be important concerning a possible quantum conjugate momentum, $\hat {\bf p}_C$, for the space variable ${\bf r}$. this can be defined as
\begin{eqnarray}
\hat {\bf p}_C =\frac{\hbar \partial}{\partial{\bf r}} =\hbar \nabla.\label{U24}
\end{eqnarray}
and this momentum exists as a result of the  Laplace equation (\ref{U23}) and automatically takes the form after operating on the wave function as follows
\begin{eqnarray}
\hat {\bf p}_C\Psi(t,\grave{\bf r})=\hbar\nabla\wedge {\bf g} (t,\grave{\bf r}),\label{U25}
\end{eqnarray}
where ${\bf g} (t,\grave{\bf r})$ is some definite vector function of  $t$ and $\grave{\bf r}$.

The wave motion followed by the dark mass dark energy time relation process can help to identify the effect that introducing position dependence has on the hyperspace vacuum. Space dependence implies the need to see this process as also space dependent.
The density functions for the dark mass, dark energy and the ratio, $r_{\Lambda,DM} (t)$, of dark energy to dark mass as functions of the time only global process are respectively represented by
\begin{eqnarray}
\rho (t) &=& (3/(8\pi G))(c/R_\Lambda)^2\sinh^{-2}(3 c t/(2R_\Lambda))\label{Cc5}\\
\rho^\dagger_\Lambda &=& (3/(4\pi G))(c/R_\Lambda)^2\label{C6}\\
r_{\Lambda,DM} (t) &=&\rho^\dagger_\Lambda /\rho (t) = 2 \sinh^{2}(3 c t/(2R_\Lambda)) \label{Cc7}\\ 
r_{\Lambda,DM} (\pm t_c) &=& 2 \sinh^{2}(\pm 3 c t_c/(2R_\Lambda))= 1.\label{Cc9}
\end{eqnarray}
The {\it space time\/} dependent version for (\ref{C5}) is given simply by multiplying both sides of this equation by the space dependant contribution $ \grave f(\grave{\bf r})\grave f^*(\grave{\bf r})$ giving
\begin{eqnarray}
\rho (t, {\bf r}) &=& (3/(8\pi G))(c/R_\Lambda)^2\grave f(\grave{\bf r})\grave f^*(\grave{\bf r})\sinh^{-2}(3 c t/(2R_\Lambda))\label{Cc10}\\
&=& (\Lambda c^2/(8\pi G)) \grave f(\grave{\bf r})\grave f^*(\grave{\bf r})\sinh^{-2}(3 c t/(2R_\Lambda)).\label{C11}
\end{eqnarray}
From equation (\ref{C11}), it follows that the cosmological constant $\Lambda$ and the space dependence function can be taken together to define a {\it local space dependent cosmological function, $\Lambda ({\bf r})$}, associated with any specific  solution of the Laplace equation as follows
\begin{eqnarray}
\Lambda ({\bf r})&=& \Lambda\grave f(\grave{\bf r})\grave f^*(\grave{\bf r})\label{Cc12}\\
&=& \Lambda\grave f({\bf r} /R_\Lambda)\grave f^*({\bf r}/R_\Lambda).\label{Cc13}
 \end{eqnarray}
It follows from this definition, that the mean value of the cosmological function is equal to $\Lambda$ for all solutions of the Laplace equation. In other words, the cosmological {\it function\/} is centred on Einstein's cosmological constant.

The linearity superposition of the various solutions of the Cosmological Schr\"odinger equation (\ref{u7}) to produce more solutions follows from (\ref{u13}) as in the following. Suppose we have two arbitrarily chosen spatially different solutions of this equation labelled with subscripts $1$ and $2$ as in
\begin{eqnarray}
\Psi_{nl,\rho ,1} (t,\grave{\bf r}) &=& \Psi_{nl,\rho ,1} (t_0,\grave{\bf r}) \exp \left(-\frac{3}{2}\int_{t_0} ^t H (t^\prime )dt^\prime \right)\nonumber\\
 \label{U26}\\
\Psi_{nl,\rho ,2} (t,\grave{\bf r}) &=& \Psi_{nl,\rho ,2} (t_0,\grave{\bf r})\exp \left(-\frac{3}{2}\int_{t_0} ^t H (t^\prime )dt^\prime \right)\nonumber\\
\label{U27}\\
\Psi_{spp}(t,\grave{\bf r})&=&c_1\Psi_{nl,\rho ,1}(t,\grave{\bf r})  + c_2\Psi_{nl,\rho ,2}(t,\grave{\bf r})\nonumber\\
&= &\Psi_{spp}(t_0, \grave{\bf r})\exp  \left(-\frac{3}{2}\int_{t_0} ^t H (t^\prime )dt^\prime \right) ,\label{U28}
\end{eqnarray}
where $c_1$ and $c_2$  are arbitrary constants and the subscript $spp$ means superposed.
It follows from (\ref{U28}) that any number of solutions of the cosmological Schr\"odiner equation can be linearly superposed to produce yet further solutions. Thus, altogether, there is vast scope to produce solutions with almost any space form whatsoever. The common feature of all the solutions is that that they all related to the common cosmological platform defined {\it by and with\/} the same time variation structure of the space constant density function of the Friedman equations. The final prescription for finding solutions to the cosmological Schrodinger equation involve the following three steps. Find any solution, $f$, to the three dimensional Laplace equation and involve in this solution one initially multiplicative arbitrary constant, $A_0$. Form the space-time wave function for this solution, $\Psi (t,{\bf r})$. Find the value of $A_0$  by using the probability normalisation condition and integration over the Hermitian square of $f$ over the volume of the universe at time, $t_c$. The wave function will then be completely determined. The probability density is also now fully determined via the definition $\rho_C(t,r) = \Psi ({t,\bf r})\Psi ^* (t, {\bf r})$. The result will be a probability density function over space and time which is compatible with the Friedman equations from general relativity. The steps will be demonstrated in the next subsection for one typical case.
\subsection{A Simple Example}
I shall finish this paper with the simplest nontrivial example giving a universe that involves a varying space and time density. One of the simplest solutions, $f(\grave{\bf r})$,  to the Laplace equation (\ref{U2.1}) is the sum of three variable complex numbers and just one arbitrary dimensionless constant, $A_0$,
\begin{eqnarray}
f(\grave{\bf r})&=&A_0((\grave x+i\grave y) +(\grave y+i\grave z)+(\grave z+i\grave x))\nonumber\\
&=& A_0(\grave x+ \grave y+\grave z)(1+i). \label{U28.1}\\
\grave f^*(\grave {\bf r})&=&  A_0(\grave x+ \grave y+\grave z)(1-i)\label{U30}\\
\grave f(\grave {\bf r})\grave f^*(\grave {\bf r})&=& 2 A_0^2(\grave x+ \grave y+\grave z)^2\label{U31.2}\\
&=& 2 (A_0/R_\Lambda)^2(x+  y+ z)^2 = F({\bf r}), \ say.\label{U31.1}
\end{eqnarray}
The definition (\ref{U31.1}) displays the formula in terms of the physical space coordinates, $x,y,z$.
The normalisation condition on the probability density, (\ref{u20.3}),  at time $t_c$ requires the following two results
\begin{eqnarray}
1& =&\frac{1}{V_U(t_c)}\int\int\int_{V_U (t_c)} \grave f(\grave{\bf r})\grave f^*(\grave{\bf r})d x dy d z\label{u20.001}\\
V_U (t_c)&=&  \left(\frac {4\pi M_U G} {3}\right)\left(\frac{R_\Lambda}{c} \right)^2.\label{u20.002} 
\end{eqnarray}
We need to evaluate the triple integral over the physical coordinates to find the value of the arbitrary constant $A_0$. This will be done in spherical polar coordinates with some condensations of notation used for the $\sin$ and $\cos$ functions,
\begin{eqnarray}
x&=& r\sin (\theta)\cos (\phi)=rS_\theta C_\phi\label{U40}\\
y&=& r\sin (\theta)\sin (\phi)=rS_\theta S_\phi\label{U41}\\
z&=& r\cos (\theta)=rC_\theta\label{U42}\\
dxdydz&=& r^2drS_\theta d\theta d\phi\label{U43.1}\\
0&<&\theta\ \le \pi,\ \ 0<\phi\le2\pi,\ \ 0<r\le r(t_c) \label{U43}\\
r(t_c)&=& (M_UG(R_\Lambda/c)^{2})^{1/3}\label{U44}
\end{eqnarray}
Thus the function, $ F({\bf r})=\grave f( \grave {\bf r})\grave f^*(\grave {\bf r})$, in the triple integral becomes
\begin{eqnarray}
F({\bf r})&=& 2 (A_0/R_\Lambda)^2(x+  y+ z)^2 \label{U45}\\
&=&2 (A_0 r/R_\Lambda)^2(S_\theta C_\phi + S_\theta S_\phi + C_\theta)^2\label{U46}\\
&=&2(A_0 r/R_\Lambda)^2((1 +2(S_\theta C_\phi S_\theta S_\phi+ S_\theta C_\phi C_\theta+ S_\theta S_\phi C_\theta))\label{U47}\\
&=&2(A_0 r/R_\Lambda)^2((1 +2(S_\theta S_\theta S_\phi C_\phi + S_\theta C_\theta C_\phi + S_\theta C_\theta S_\phi)) \label{U48}
\end{eqnarray}
Introducing the further notation
\begin{eqnarray}
i_r&=& r^4 dr\ \ \ \ \ \ \ \ I_r= \int_0^{r(t_c)} i_r=r^5(t_c)/5\label{U48.1}\\
i_{\theta,1} &=& S_\theta^3 d\theta ,\ \ \ \ \ \  I_1= \int_0^\pi i_{\theta,1}=\frac{4}{3} \label{U49}\\
i_{\phi,1} &=& S_\phi C_\phi d\phi ,\ \  I_2= \int_0^{2\pi} i_{\phi,1}=0 \label{U50}\\
i_{\theta,2}&=& S_\theta^2 C_\theta d\theta,\ \  I_3= \int_0^\pi i_{\theta,2} =0 \label{U51}\\
i_{\phi,2} &=& C_\phi d\phi,\ \ \ \  I_4= \int_0^{2\pi} i_{\phi,2} =0 \label{U52}\\
i_{\phi,3} &=& S_\phi d\phi, \ \ \ \ I_5= \int_0^{2\pi} i_{\phi,3} =0 \label{U70}
\end{eqnarray}
the integral element and the integral can be expressed as
\begin{eqnarray}
dI &=&2(A_0 r/R_\Lambda)^2((1 +2(S_\theta S_\theta S_\phi C_\phi + S_\theta C_\theta C_\phi + S_\theta C_\theta S_\phi)) r^2drS_\theta d\theta d\phi\nonumber\\ 
&=&  2(A_0 /R_\Lambda)^2i_r (d\theta d\phi +2(i_{\theta,1} i_{\phi,1}+ i_{\theta,2} i_{\phi,2} + i_{\theta,2} i_{\phi,3}))\label{U72}
\end{eqnarray}
\begin{eqnarray}
I(t_c)&=&  (4/5) \left(\frac{A_0 \pi }{R_\Lambda}\right)^2 r^5(t_c).\label{U71}
\end{eqnarray}
The last expression for $I(t_c)$ is all that is left after integration.
The normalisation condition at time $t_c$ using (\ref{u20.4}) can now be used to find the numerical value of $A_0$ by
\begin{eqnarray}
1&=&I/V_U(t_c)= (4/5) \left(\frac{A_0 \pi }{R_\Lambda^2}\right)^2 r^5(t_c)\left(\frac {3 c^2} {4\pi M_U G }\right)\label{U73}\\
&=& \frac{3 A_0^2\pi}{5}\left(\frac{M_UG}{c^2R_\Lambda}\right)^{2/3}\\
A_0&=& \left( \frac{5}{3\pi}\right)^{1/2}\left( \frac {c^2R_\Lambda}{M_UG}   \right)^{1/3} \label{U75}
\end{eqnarray}
Thus the full solution for the wave function, the probability density and all the constants involved is as follows:
\begin{eqnarray}
\Psi_{nl,\rho} (t,\grave{\bf r}) &=& \Psi_{nl,\rho} (t_c,\grave{\bf r}) \exp \left(-\frac{3}{2}\int_{t_c} ^t H (t^\prime )dt^\prime \right) \label{U76}\\
 \Psi_{nl,\rho} (t_c,\grave{\bf r})&=& \Psi_{nl,\rho}(t_c) \grave f(\grave{\bf r}) \label{U77}\\
\Psi_{nl,\rho}(t_c) &=& ( \rho^\dagger_\Lambda)^{1/2} \label{U78}\\
\grave f(\grave{\bf r})&=& (A_0/R_\Lambda )(x+y+z)(1+i) \label{U79}\\
\rho (t, {\bf r})&=&2 (A_0/R_\Lambda)^2\rho^\dagger_\Lambda (x+y+z)^2\exp \left(-3\int_{t_c} ^t H (t^\prime )dt^\prime \right)\label{U80}\\
\rho^\dagger_\Lambda &=& \frac{\Lambda c^2}{4\pi G}\label{U82}\\
A_0&=& \left( \frac{5}{3\pi}\right)^{1/2}\left( \frac {c^2R_\Lambda}{M_UG}   \right)^{1/3}.\label{U84}
\end{eqnarray}
\section{Appendix 4 Conclusions}
In an earlier paper, it was shown that a non-linear Shr\"odinger equation can be obtained from the Friedman cosmology equations which is entirely consistent with those equations. Here, the time evolution of this Schr\"odinger equation is examined in relation to conservation of the universe's total positive gravitational mass. This leads to the identification of a wave function for cosmology states with a definite time evolution and consequently also to a probability density for cosmology. This cosmological probability density  can depend on spatial variability in addition to just the time variability of the Friedman equation structure. Consistency of the new Schr\"odinger equation with its originating Friedman set is achieved by restricting solutions to the condition that they satisfy the Laplace equation in hyperspace. It becomes clear that, even with this restriction, a multiple infinity of solutions remain available and applicable.    
The structure of this theory seems to confirm the view often expressed about the {\it quantum vacuum\/} that it is a bubbling cauldron of activity in the form of random quantum transitions, such as pair production and annihilation, between short lived virtual states of fundamental particles. The expansion of the universe can be explained in such terms as a spherical advancing and evolving wave of quantum {\it before and after measurement type conditions\/} in reverse through the expanding boundary. Just outside the expanding boundary, the vacuum chaotic states as described by the {\it wave function\/}, resourced by the multiplicity of solutions of the Laplace equation, are progressively converted from chaos to a definite  gravitational form sufficient to describe the mass density that has taken up residence  within the expanded boundary.  The universe expansion colonises surrounding hyperspace so as to accommodate within its boundary its {\it conserved positive\/} gravitational mass with more territory and in a quantum form that can hold non-transient positive gravitational mass. Outside the universe the solution holds but remains a linear superposition of many varied chaotic transient states with mass density value centred on the value of twice Einstein's dark energy mass density $\rho^\dagger_\Lambda$.      


\begin{thebibliography}{99}
\bibitem{01:kmo}{R. A. Knop et al. arxiv.org/abs/astro-ph/0309368\\
\href{http://arxiv.org/abs/astro-ph/0309368}{New Constraints}
on $\Omega_M$, $\Omega_\Lambda$ and $\omega$ from\\
an independent Set (Hubble) of Eleven High-Redshift\\
Supernovae, Observed with HST}
\bibitem{02:rie}{Adam G. Riess et al xxx.lanl.gov/abs/astro-ph/0402512\\
\href{http://xxx.lanl.gov/abs/astro-ph/0402512}{Type 1a Supernovae Discoveries} at $z>1$\\ 
From The Hubble Space Telescope: Evidence for Past\\
Deceleration and constraints on Dark energy Evolution}
\bibitem{44:berr}{Berry 1978, Principles of cosmology and gravitation, CUP}
\bibitem{04:gil}{Gilson, J.G. 1991, Oscillations of a Polarizable Vacuum,\\ Journal of Applied Mathematics and Stochastic Analysis,\\
{\bf 4}, 11, 95--110.}
\bibitem{05:gil}{Gilson, J.G. 1994, Vacuum Polarisation and\\
The Fine Structure Constant, Speculations in Science\\
and Technology , {\bf  17}, 3 , 201-204.}
\bibitem{06:gil}{Gilson, J.G. 1996, Calculating the fine structure constant,\\
Physics Essays, {\bf  9} , 2 June, 342-353.}
\bibitem{07:edd}{Eddington, A.S. 1946, Fundamental Theory, Cambridge\\
University Press.}
\bibitem{08:kil}{Kilmister, C.W. 1992, Philosophica, {\bf  50}, 55.} \bibitem{09:bas}{Bastin, T., Kilmister, C. W. 1995, Combinatorial Physics\\ World Scientific Ltd.}
\bibitem{10:kil}{Kilmister, C. W. 1994 , Eddington's search for a Fundamental\\ Theory, CUP.}
\bibitem{18:moh}{Peter, J. Mohr, Barry, N. Taylor, 1998,\\
 Recommended Values of the fundamental Physical  Constants,\\
Journal of Physical and Chemical Reference Data, AIP}
\bibitem{19:gil}{Gilson,  J. G. 1997, Relativistic Wave Packing\\
and  Quantization, Speculations in Science and Technology,\\
{\bf  20} Number 1, March, 21-31}
\bibitem{28:dir}{Dirac, P. A. M.  1931, Proc. R. Soc. London, {\bf  A133}, 60.}
\bibitem{32:gil}{ Gilson, J.G. 2007, www.fine-structure-constant.org\\ \href{http://www.fine-structure-constant.org/}{The fine structure constant}}
\bibitem{33:mcp}{McPherson R., Stoney Scale and Large Number\\
Coincidences, Apeiron, Vol. 14, No. 3, July, 2007}
\bibitem{03:rind} {Rindler, W. 2006, Relativity: Special, General\\
and Cosmological, Second Edition, Oxford University Press}
\bibitem{3:mis}{Misner, C. W.; Thorne, K. S.; and Wheeler, J. A. 1973,\\ Gravitation, Boston, San Francisco, CA: W. H. Freeman}
\bibitem{39:gil}{J. G. Gilson, 2004, Physical Interpretations of\\
Relativity Theory Conference IX\\
London, Imperial College, September, 2004\\
\href{http://arxiv.org/abs/physics/0409010}
{Mach's Principle II}}
\bibitem{40:gil}{J. G. Gilson, A Sketch for a Quantum Theory of Gravity:\\
Rest Mass Induced by Graviton Motion, May/June 2006,\\
Vol. 17, No. 3, Galilean Electrodynamics}
\bibitem{41:gil}{J. G. Gilson, arxiv.org/PS$\_$cache/physics/pdf/0411/0411085v2.pdf\\
\href{http://arxiv.org/PS_cache/physics/pdf/0411/0411085v2.pdf}{A Sketch for a Quantum Theory of Gravity:}\\
Rest Mass Induced by Graviton Motion}
\bibitem{42:gil}{J. G. Gilson, arxiv.org/PS$\_$cache/physics/pdf/0504/0504106v1.pdf\\
\href{http://arxiv.org/PS_cache/physics/pdf/0504/0504106v1.pdf}{Dirac's Large Number Hypothesis}\\
and Quantized Friedman Cosmologies}
\bibitem{43:nar}{Narlikar, J. V., 1993, Introduction to Cosmology, CUP}
\bibitem{45:gil}{Gilson, J.G. 2005, A Dust Universe Solution to the Dark Energy\\ Problem, Vol. 1, {\it Aether, Spacetime and Cosmology\/},\\
PIRT publications, 2007,\\ \href{http://arxiv.org/PS\_cache/physics/pdf/0512/0512166v2.pdf}{ arxiv.org/PS\_cache/physics/pdf/0512/0512166v2.pdf }}
\bibitem{46:gil}{Gilson, PIRT Conference 2006, Existence of Negative Gravity\\ Material, Identification of Dark Energy,\\
\href{http://arxiv.org/abs/physics/0603226}{ arxiv.org/abs/physics/0603226}}
\bibitem{47:lem}{G.  Lema\^itre, Ann. Soc. Sci. de Bruxelles\\
Vol. A47, 49, 1927}
\bibitem{48:adl}{Ronald J. Adler, James D. Bjorken and James M. Overduin 2005,\\
\href{http://www.arxiv.org/abs/gr-qc/0602102}{Finite cosmology and a CMB cold spot, SLAC-PUB-11778}}
\bibitem{49:man}{Mandl, F., 1980, Statistical Physics, John Wiley}
\bibitem{54:riz}{Rizvi 2005, Lecture 25, PHY-302,\\ \href{http://hepwww.ph.qmw.ac.uk/~rizvi/npa/NPA-25.pdf}{http://hepwww.ph.qmw.ac.uk/$\sim$rizvi/npa/NPA-25.pdf}}
\bibitem{52:ham}{Nicolay J. Hammer, 2006\\
\href{http://www.mpa-garching.mpg.de/lectures/ADSEM/SS06_Hammer.pdf}{www.mpa-garching.mpg.de/lectures/ADSEM/SS06\_Hammer.pdf}}
\bibitem{53:pap}{E. M. Purcell, R. V. Pound, 1951, Phys. Rev.,{\bf 81, 279}}
\bibitem{55:gil}{Gilson J. G., 2006, www.maths.qmul.ac.uk/$\sim$ jgg/darkenergy.pdf\\ \href{http://www.maths.qmul.ac.uk/~jgg/darkenergy.pdf}{Presentation to PIRT Conference 2006}}
\bibitem{56:gil}{Gilson J. G., 2007, Thermodynamics of a Dust Universe,\\Energy density, Temperature, Pressure and Entropy\\for Cosmic Microwave Background}{\href{http://arxiv.org/abs/physics/0701286}{
http://arxiv.org/abs/0704.2998}}
\bibitem{57:bec}{Beck, C., Mackey, M. C. \href{http://xxx.arxiv.org/abs/astro-ph/0406504}{http://xxx.arxiv.org/abs/astro-ph/0406504}}
\bibitem{58:gil}{Gilson J. G., 2007, Reconciliation of Zero-Point and Dark Energies \\in a Friedman Dust Universe with\\ Einstein's Lambda,
}{\href{http://arxiv.org/abs/0704.2998}{
http://arxiv.org/abs/0704.2998}}
\bibitem{59:rud}{Rudnick L. et al, 2007, \href{http://arxiv.org/abs/0704.0908v2}
{WMP Cold Spot, Apj in press}}
\bibitem{60:gil}{Gilson J. G., 2007, Cosmological Coincidence Problem in\\
an Einstein Universe and in a Friedman Dust Universe with\\ Einstein's Lambda, }{Vol. 2, {\href{http://arxiv.org/abs/0705.2872}
{\it Aether, Spacetime and Cosmology\/},\\ PIRT publications, 2008   }}
\bibitem{61:free}{Freedman W. L. and Turner N. S., 2008, Observatories of the Carnegie Institute Washington,
Measuring and Understanding the Universe}
\bibitem{62:gil}{Gilson J. G., 2007, Expanding Boundary Pressure Process.\\All pervading Dark Energy Aether in a Friedman Dust Universe \\
with Einstein's Lambda, }{Vol. 2, { \href{http://www.maths.qmul.ac.uk/~jgg/gil110.pdf} 
{\it Aether, Spacetime and Cosmology\/},\\ PIRT publications, 2008}}
\bibitem{63:gil}{Gilson J. G., 2007, Fundamental Dark Mass, Dark Energy Time\\ Relation in a Friedman Dust Universe and in a Newtonian Universe\\
with Einstein's Lambda, }{Vol. 2, { \href{http://www.maths.qmul.ac.uk/~jgg/gil108.pdf} 
{\it Aether, Spacetime and Cosmology\/},\\ PIRT publications, 2008}}
\bibitem{64:gil}{Gilson J. G., 2008, . A quantum Theory Friendly Cosmology\\Exact Gravitational Waves in a Friedman Dust Universe \\
with Einstein's Lambda,}
\href{http://www.maths.qmul.ac.uk/~jgg/gil111.pdf} 
{PIRT Conference, 2008}
\end{thebibliography}
 \end{document}